




\catcode`\@=13                                                         
\def@{\errmessage{AmS-TeX error: \string@ has no current use
     (use \string\@\space for printed \string@ symbol)}}
\catcode`\@=11                                                         
\def\@{\char'100 }
\catcode`\~=13                                                         


\def\err@AmS#1{\errmessage{AmS-TeX error: #1}}                         


\def\eat@AmS#1{}

\long\def\comp@AmS#1#2{\def\@AmS{#1}\def\@@AmS{#2}\ifx
   \@AmS\@@AmS\def\cresult@AmS{T}\else\def\cresult@AmS{F}\fi}          

\def\in@AmS#1#2{\def\intest@AmS##1#1##2{\comp@AmS##2\end@AmS\if T\cresult@AmS
   \def\cresult@AmS{F}\def\in@@AmS{}\else
   \def\cresult@AmS{T}\def\in@@AmS####1\end@AmS{}\fi\in@@AmS}%
   \def\cresult@AmS{F}\intest@AmS#2#1\end@AmS}                         


\let\relax@AmS=\relax                                                  


\def\magstep#1{\ifcase#1 \@m\or 1200\or 1440\or 1728\or 2074\or 2488\fi
     \relax@AmS}

\def\iterate{\body\let\next\iterate \else\let\next\relax@AmS\fi \next}

\def\enskip{\hskip.5em\relax@AmS}

\def\strut{\relax@AmS\ifmmode\copy\strutbox\else\unhcopy\strutbox\fi}

\let\+=\relax@AmS
\def\sett@b{\ifx\next\+\let\next\relax@AmS
    \def\next{\afterassignment\s@tt@b\let\next}%
  \else\let\next\s@tcols\fi\next}
\def\s@tt@b{\let\next\relax@AmS\us@false\m@ketabbox}

\def\smash{\relax@AmS 
  \ifmmode\def\next{\mathpalette\mathsm@sh}\else\let\next\makesm@sh
  \fi\next}


\def\define#1{\expandafter\ifx\csname\expandafter\eat@AmS\string#1\endcsname
   \relax@AmS\def\dresult@AmS{\def#1}\else
   \err@AmS{\string#1\space is already defined}\def
      \dresult@AmS{\def\garbage@AmS}\fi\dresult@AmS}                   

\def\predefine#1#2{\let#1=#2}


\chardef\plus=`+
\chardef\equal=`=
\chardef\less=`<
\chardef\more=`>


\let\ic@AmS=\/
\def\/{\unskip\ic@AmS}

\def\Space@AmS.{\futurelet\Space@AmS\relax@AmS}
\Space@AmS. 

\def~{\unskip\futurelet\tok@AmS\s@AmS}
\def\s@AmS{\ifx\tok@AmS\Space@AmS\def\next@AmS{}\else
        \def\next@AmS{\ }\fi\penalty 9999 \next@AmS}                  

\def\period{\unskip.\spacefactor3000 { }}

\def\srdr@AmS{\thinspace}                                             
\def\drsr@AmS{\kern .02778em }
\def\sldl@AmS{\kern .02778em}
\def\dlsl@AmS{\thinspace}

\def\lqtest@AmS#1{\comp@AmS{#1}`\if T\cresult@AmS\else\comp@AmS{#1}\lq\fi}


\def\qspace#1{\unskip
  \lqtest@AmS{#1}\let\fresult@AmS=\cresult@AmS\if T\cresult@AmS
     \def\qspace@AmS{\ifx\tok@AmS\Space@AmS\def\next@AmS{\dlsl@AmS`}\else
       \def\next@AmS{\qspace@@AmS}\fi\next@AmS}\else
     \def\qspace@AmS{\ifx\tok@AmS\Space@AmS\def\next@AmS{\drsr@AmS'}\else
       \def\next@AmS{\qspace@@AmS}\fi\next@AmS}\fi
    \futurelet\tok@AmS\qspace@AmS}                                    

\def\qspace@@AmS{\futurelet\tok@AmS\qspace@@@AmS}

\def\qspace@@@AmS{\if T\fresult@AmS  \ifx\tok@AmS`\sldl@AmS`\else
       \ifx\tok@AmS\lq\sldl@AmS`\else \dlsl@AmS`\fi \fi
                         \else  \ifx\tok@AmS'\srdr@AmS'\else
        \ifx\tok@AmS\rq\srdr@AmS'\else \drsr@AmS'\fi \fi
        \fi}

\def\{{\relax@AmS\ifmmode\delimiter"4266308 \else
    $\delimiter"4266308 $\fi}                            

\def\}{\relax@AmS\ifmmode\delimiter"5267309 \else$\delimiter"5267309 $\fi}

\def\AmSTeX{$\cal A$\kern-.1667em\lower.5ex\hbox{$\cal M$}\kern-.125em
     $\cal S$-\TeX}


\def\linebreak{\unskip\penalty-10000 }                                
\def\pagebreak{\vadjust{\penalty-10000 }}

\def\newline{\ifvmode \err@AmS{There's no line here to break}\else
     \hfil\penalty-10000 \fi}

\def\topspace#1{\insert\topins{\penalty100 \splittopskip=0pt
     \vbox to #1{}}}
\def\midspace#1{\setbox0=\vbox to #1{}\advance\dimen0 by \pagetotal
  \ifdim\dimen0>\pagegoal\topspace{#1}\else\vadjust{\box0}\fi}

\long\def\comment{\begingroup
 \catcode`\{=12 \catcode`\}=12 \catcode`\#= 12 \catcode`\^^M=12
   \catcode`\%=12 \catcode`^^A=14
    \comment@AmS}
\begingroup\catcode`^^A=14
\catcode`\^^M=12  ^^A
\long\gdef\comment@AmS#1^^M#2{\comp@AmS\endcomment{#2}\if T\cresult@AmS^^A
\def\comment@@AmS{\endgroup}\else^^A
 \long\def\comment@@AmS{\comment@AmS#2}\fi\comment@@AmS}\endgroup     


\def\text#1{\hbox{\rm#1}}

\def\quad{\relax@AmS\ifmmode
    \hbox{\hskip1em}\else\hskip1em\relax@AmS\fi}                      
\def\qquad{\quad\quad}
\def\,{\relax@AmS\ifmmode\mskip\thinmuskip\else$\mskip\thinmuskip$\fi}
\def\;{\relax@AmS
  \ifmmode\mskip\thickmuskip\else$\mskip\thickmuskip$\fi}

\def\frac#1#2{{#1\over#2}}

\mathchardef\:="603A                                                  


\def\big@AmS#1{{\hbox{$\left#1\vbox to\big@@AmS{}\right.\offspace@AmS$}}}
\def\Big@AmS#1{{\hbox{$\left#1\vbox to\Big@@AmS{}\right.\offspace@AmS$}}}
\def\bigg@AmS#1{{\hbox{$\left#1\vbox to\bigg@@AmS{}\right.\offspace@AmS$}}}
\def\Bigg@AmS#1{{\hbox{$\left#1\vbox to\Bigg@@AmS{}\right.\offspace@AmS$}}}
\def\offspace@AmS{\nulldelimiterspace0pt \mathsurround0pt }

\def\big@@AmS{8.5pt}                                
\def\Big@@AmS{11.5pt}
\def\bigg@@AmS{14.5pt}
\def\Bigg@@AmS{17.5pt}

\def\bigl{\mathopen\big@AmS}
\def\bigm{\mathrel\big@AmS}
\def\bigr{\mathclose\big@AmS}
\def\Bigl{\mathopen\Big@AmS}
\def\Bigm{\mathrel\Big@AmS}
\def\Bigr{\mathclose\Big@AmS}
\def\biggl{\mathopen\bigg@AmS}
\def\biggm{\mathrel\bigg@AMS}
\def\biggr{\mathclose\bigg@AmS}
\def\Biggl{\mathopen\Bigg@AmS}
\def\Biggm{\mathrel\Bigg@AmS}
\def\Biggr{\mathclose\Bigg@AmS}


{\catcode`'=13 \gdef'{^\bgroup\prime\prime@AmS}}
\def\prime@AmS{\futurelet\tok@AmS\prime@@AmS}
\def\prime@@@AmS#1{\futurelet\tok@AmS\prime@@AmS}
\def\prime@@AmS{\ifx\tok@AmS'\def\next@AmS{\prime\prime@@@AmS}\else
   \def\next@AmS{\egroup}\fi\next@AmS}


\def\topsmash{\relax@AmS\ifmmode\def\topsmash@AmS
   {\mathpalette\mathtopsmash@AmS}\else
    \let\topsmash@AmS=\maketopsmash@AmS\fi\topsmash@AmS}

\def\maketopsmash@AmS#1{\setbox0=\hbox{#1}\topsmash@@AmS}

\def\mathtopsmash@AmS#1#2{\setbox0=\hbox{$#1{#2}$}\topsmash@@AmS}

\def\topsmash@@AmS{\vbox to 0pt{\kern-\ht0\box0}}

\def\botsmash{\relax@AmS\ifmmode\def\botsmash@AmS
   {\mathpalette\mathbotsmash@AmS}\else
     \let\botsmash@AmS=\makebotsmash@AmS\fi\botsmash@AmS}

\def\makebotsmash@AmS#1{\setbox0=\hbox{#1}\botsmash@@AmS}

\def\mathbotsmash@AmS#1#2{\setbox0=\hbox{$#1{#2}$}\botsmash@@AmS}

\def\botsmash@@AmS{\vbox to \ht0{\box0\vss}}


\def\LimitsOnSums{\let\slimits@AmS=\displaylimits}                    
\def\NoLimitsOnSums{\let\slimits@AmS=\nolimits}

\LimitsOnSums

\mathchardef\coprod@AmS"1360       \def\coprod{\coprod@AmS\slimits@AmS}
\mathchardef\bigvee@AmS"1357       \def\bigvee{\bigvee@AmS\slimits@AmS}
\mathchardef\bigwedge@AmS"1356     \def\bigwedge{\bigwedge@AmS\slimits@AmS}
\mathchardef\biguplus@AmS"1355     \def\biguplus{\biguplus@AmS\slimits@AmS}
\mathchardef\bigcap@AmS"1354       \def\bigcap{\bigcap@AmS\slimits@AmS}
\mathchardef\bigcup@AmS"1353       \def\bigcup{\bigcup@AmS\slimits@AmS}
\mathchardef\prod@AmS"1351         \def\prod{\prod@AmS\slimits@AmS}
\mathchardef\sum@AmS"1350          \def\sum{\sum@AmS\slimits@AmS}
\mathchardef\bigotimes@AmS"134E    \def\bigotimes{\bigotimes@AmS\slimits@AmS}
\mathchardef\bigoplus@AmS"134C     \def\bigoplus{\bigoplus@AmS\slimits@AmS}
\mathchardef\bigodot@AmS"134A      \def\bigodot{\bigodot@AmS\slimits@AmS}
\mathchardef\bigsqcup@AmS"1346     \def\bigsqcup{\bigsqcup@AmS\slimits@AmS}

\def\LimitsOnInts{\let\ilimits@AmS=\displaylimits}
\def\NoLimitsOnInts{\let\ilimits@AmS=\nolimits}

\NoLimitsOnInts

\mathchardef\int@AmS"1352
\def\int{\gdef\intflag@AmS{T}\int@AmS\ilimits@AmS}                    

\mathchardef\oint@AmS"1348 \def\oint{\gdef\intflag@AmS{T}\oint@AmS\ilimits@AmS}

\def\inttest@AmS#1{\def\intflag@AmS{F}\setbox0=\hbox{$#1$}}

\def\intic@AmS{\mathchoice{\hbox{\hskip5pt}}{\hbox
          {\hskip4pt}}{\hbox{\hskip4pt}}{\hbox{\hskip4pt}}}           
\def\negintic@AmS{\mathchoice
  {\hbox{\hskip-5pt}}{\hbox{\hskip-4pt}}{\hbox{\hskip-4pt}}{\hbox{\hskip-4pt}}}
\def\intkern@AmS{\mathchoice{\!\!\!}{\!\!}{\!\!}{\!\!}}
\def\intdots@AmS{\mathchoice{\cdots}{{\cdotp}\mkern 1.5mu
    {\cdotp}\mkern 1.5mu{\cdotp}}{{\cdotp}\mkern 1mu{\cdotp}\mkern 1mu
      {\cdotp}}{{\cdotp}\mkern 1mu{\cdotp}\mkern 1mu{\cdotp}}}

\newcount\intno@AmS                                                   

\def\intii{\gdef\intflag@AmS{T}\intno@AmS=2\futurelet                 
              \tok@AmS\ints@AmS}
\def\intiii{\gdef\intflag@AmS{T}\intno@AmS=3\futurelet\tok@AmS\ints@AmS}
\def\intiv{\gdef\intflag@AmS{T}\intno@AmS=4\futurelet\tok@AmS\ints@AmS}
\def\intdotsint{\gdef\intflag@AmS{T}\intno@AmS=0\futurelet
    \tok@AmS\ints@AmS}

\def\ints@AmS{\findlimits@AmS\ints@@AmS}

\def\findlimits@AmS{\def\ignoretoken@AmS{T}\ifx\tok@AmS\limits
   \def\limits@AmS{T}\else\ifx\tok@AmS\nolimits\def\limits@AmS{F}\else
     \def\ignoretoken@AmS{F}\ifx\ilimits@AmS\nolimits\def\limits@AmS{F}\else
       \def\limits@AmS{T}\fi\fi\fi}

\def\multintlimits@AmS{\int@AmS\ifnum \intno@AmS=0\intdots@AmS
  \else \intkern@AmS\fi
    \ifnum\intno@AmS>2\int@AmS\intkern@AmS\fi
     \ifnum\intno@AmS>3 \int@AmS\intkern@AmS\fi \int@AmS}

\def\multint@AmS{\int\ifnum \intno@AmS=0\intdots@AmS\else\intkern@AmS\fi
   \ifnum\intno@AmS>2\int\intkern@AmS\fi
    \ifnum\intno@AmS>3 \int\intkern@AmS\fi \int}

\def\ints@@AmS{\if F\ignoretoken@AmS\def\ints@@@AmS{\if
    T\limits@AmS\negintic@AmS
 \mathop{\intic@AmS\multintlimits@AmS}\limits\else
    \multint@AmS\nolimits\fi}\else\def\ints@@@AmS{\if T\limits@AmS
   \negintic@AmS\mathop{\intic@AmS\multintlimits@AmS}\limits\else
    \multint@AmS\nolimits\fi\eat@AmS}\fi\ints@@@AmS}

\def\LimitsOnNames{\let\nlimits@AmS=\displaylimits}
\def\NoLimitsOnNames{\let\nlimits@AmS=\nolimits}

\LimitsOnNames


\def\operatornamewithlimits#1{\mathop{\mathcode`'="7027 \mathcode`-="702D
   \rm #1}\nlimits@AmS}

\def\liminj{\setbox0=\hbox{\rm lim}\mathop{\rm lim}
		\limits_{\topsmash{\hbox to \wd0{\leftarrowfill}}}}
\def\limproj{\setbox0=\hbox{\rm lim}\mathop{\rm lim}
		\limits_{\topsmash{\hbox to \wd0{\rightarrowfill}}}}


\newdimen\buffer@AmS
\buffer@AmS=\fontdimen13\tenex                                        
\newdimen\buffer
\buffer=\buffer@AmS

\def\resetbuffer{\fontdimen13 \tenex=\buffer@AmS \buffer=\buffer@AmS}


\def\Let@AmS{\relax@AmS\iffalse{\fi\let\\=\cr\iffalse}\fi}            

\def\align{\def\vspace##1{\noalign{\vskip ##1}}                       
  \,\vcenter\bgroup\Let@AmS\tabskip=0pt\openup3pt\mathsurround=0pt  
  \halign\bgroup\strut
  \hfil$\displaystyle{##}$&$\displaystyle{{}##}$\hfil\cr}        

\def\endalign{\strut\crcr\egroup\egroup}

\def\bunch{\def\vspace##1{\noalign{\vskip ##1}}
  \,\vcenter\bgroup\Let@AmS\tabskip=0pt\openup3pt\mathsurround=0pt
     \halign\bgroup\strut\hfil$\displaystyle{##}$\hfil\cr}

\def\endbunch{\strut\crcr\egroup\egroup}

\def\matrix{\catcode`\^^I=4 \futurelet\tok@AmS\matrix@AmS}            

\def\matrix@AmS{\relax@AmS\ifnum`}=0\fi\ifx\tok@AmS\format
   \def\next@AmS{\expandafter\matrix@@AmS\eat@AmS}\else
   \def\next@AmS{\matrix@@@AmS}\fi\next@AmS}

\def\matrix@@@AmS{
 \ifnum`{=0\fi\iffalse}\fi\,\vcenter\bgroup\Let@AmS\tabskip=0pt
    \normalbaselines\halign\bgroup $\strut\hfil##\hfil$&&\quad$\strut
  \hfil##\hfil$\cr\strut\cr\noalign{\kern-\baselineskip}}             

\def\matrix@@AmS#1\\{
   \def\premable@AmS{#1}\toks@{##}
 \def\c{$\copy\strutbox\hfil\the\toks@\hfil$}\def\r
   {$\copy\strutbox\hfil\the\toks@$}%
   \def\l{$\copy\strutbox\the\toks@\hfil$}%
\setbox0=
\hbox{\xdef\Preamble@AmS{\premable@AmS}}
 \def\vspace##1{\noalign{\vskip ##1}}\ifnum`{=0\fi\iffalse}\fi
\,\vcenter\bgroup\Let@AmS
  \tabskip=0pt\normalbaselines\halign\bgroup\span\Preamble@AmS\cr
    \mathstrut\cr\noalign{\kern-\baselineskip}}

\def\endmatrix{\crcr\mathstrut\cr\noalign{\kern-\baselineskip
   }\egroup\egroup\,\catcode`\^^I=10 }

\def\spacedots#1for#2{\multispan#2\leaders\hbox{$\mkern#1mu.\mkern
    #1mu$}\hfill}

\def\enabletabs{\catcode`\^^I=4 \enabletabs@AmS}
\def\enabletabs@AmS#1\disabletabs{#1\catcode`\^^I=10 }                

\def\smallmatrix{\futurelet\tok@AmS\smallmatrix@AmS}                  

\def\smallmatrix@AmS{\relax@AmS\ifnum`}=0\fi\ifx\tok@AmS\format
   \def\next@AmS{\expandafter\smallmatrix@@AmS\eat@AmS}\else
   \def\next@AmS{\smallmatrix@@@AmS}\fi\next@AmS}

\def\smallmatrix@@@AmS{
 \ifnum`{=0\fi\iffalse}\fi\,\vcenter\bgroup\Let@AmS\tabskip=0pt
    \baselineskip8pt\lineskip1pt\lineskiplimit1pt
  \halign\bgroup $\strut\hfil##\hfil$&&\;$\strut
  \hfil##\hfil$\cr\strut\cr\noalign{\kern-\baselineskip}}

\def\smallmatrix@@AmS#1\\{
   \def\premable@AmS{#1}\toks@{##}
 \def\c{$\copy\strutbox\hfil\the\toks@\hfil$}\def\r
   {$\copy\strutbox\hfil\the\toks@$}%
   \def\l{$\copy\strutbox\the\toks@\hfil$}%
\hbox{\xdef\Preamble@AmS{\premable@AmS}}
 \def\vspace##1{\noalign{\vskip ##1}}\ifnum`{=0\fi\iffalse}\fi
\,\vcenter\bgroup\Let@AmS
     \tabskip=0pt\baselineskip8pt\lineskip1pt\lineskiplimit1pt
\halign\bgroup\span\Preamble@AmS\cr
    \mathstrut\cr\noalign{\kern-\baselineskip}}

\def\endsmallmatrix{\crcr\mathstrut\cr\noalign{\kern-\baselineskip}
   \egroup\egroup\,}

\def\cases{\left\{ \,\vcenter\bgroup\Let@AmS\normalbaselines\tabskip=0pt
   \halign\bgroup$##\hfil$&\qquad$##\hfil$\cr}                        

\def\endcases{\crcr\egroup\egroup\right.}


\def\TagsOnLeft{\def\tagposition@AmS{L}}
\def\TagsOnRight{\def\tagposition@AmS{R}}
\def\TagsAsMath{\def\tagstyle@AmS{M}}
\def\TagsAsText{\def\tagstyle@AmS{T}}

\TagsOnLeft
\TagsAsText

\def\tag#1$${\if L\tagposition@AmS
    \leqno\else\eqno\fi\def\atag@AmS{T}\maketag@AmS#1\tagend@AmS$$}   

\def\maketag@AmS{\futurelet\tok@AmS\maketag@@AmS}                     
\def\maketag@@AmS{\ifx\tok@AmS[\def\next@AmS{\maketag@@@AmS}\else
      \def\next@AmS{\maketag@@@@AmS}\fi\next@AmS}
\def\maketag@@@AmS[#1]#2\tagend@AmS{\if F\atag@AmS\else             
   \if M\tagstyle@AmS\hbox{$#1$}\else\hbox{#1}\fi\fi
       \gdef\atag@AmS{F}}
\def\maketag@@@@AmS#1\tagend@AmS{\if F\atag@AmS \else
        \if T\autotag@AmS \setbox0=\hbox
    {\if M\tagstyle@AmS\tagform@AmS{$#1$}\else\tagform@AmS{#1}\fi}
                        \ifdim\wd0=0pt \tagform@AmS{*}\else
            \if M\tagstyle@AmS\tagform@AmS{$#1$}\else\tagform@AmS{#1}\fi
                     \fi\else
               \if M\tagstyle@AmS\tagform@AmS{$#1$}\else\tagform@AmS{#1}\fi
                     \fi
                  \fi\gdef\atag@AmS{F}}

\def\tagform@AmS#1{\hbox{\rm(#1\unskip)}}

\def\AutoTag{\def\autotag@AmS{T}}
\def\NoAutoTag{\def\autotag@AmS{F}}

\NoAutoTag

\def\inaligntag@AmS{F} \def\inbunchtag@AmS{F}                         

\def\CenteredTagsOnBrokens{\def\centerbroken@AmS{T}}                  
\def\TopOrBottomTagsOnBrokens{\def\centerbroken@AmS{F}}
\TopOrBottomTagsOnBrokens

\def\broken{\global\setbox0=\vbox\bgroup\Let@AmS\tabskip=0pt
 \if T\inaligntag@AmS\else
   \if T\inbunchtag@AmS\else\openup3pt\fi\fi\mathsurround=0pt
     \halign\bgroup\strut\hfil$\displaystyle{##}$&$\displaystyle{{}##}$\hfill
      \cr}
\def\endbroken{\strut\crcr\egroup\egroup
      \global\setbox7=\vbox{\unvbox0\setbox1=\lastbox
      \hbox{\unhbox1\unskip\setbox2=\lastbox
       \unskip\setbox3=\lastbox
         \global\setbox4=\copy3
          \box3\box2}}
  \if L\tagposition@AmS
     \if T\inaligntag@AmS
           \if T\centerbroken@AmS\gdef\broken@AmS
                {&\vcenter{\vbox{\moveleft\wd4\box7}}}
           \else
            \gdef\broken@AmS{&\vbox{\moveleft\wd4\vtop{\unvbox7}}}
           \fi
     \else                                                            
           \if T\centerbroken@AmS\gdef\broken@AmS
                {\vcenter{\box7}}%
           \else
              \gdef\broken@AmS{\vtop{\unvbox7}}%
           \fi
     \fi
  \else                                                  
      \if T\inaligntag@AmS
           \if T\centerbroken@AmS
              \gdef\broken@AmS{&\vcenter{\vbox{\moveleft\wd4\box7}}}%
          \else
             \gdef\broken@AmS{&\vbox{\moveleft\wd4\box7}}%
          \fi
      \else
          \if T\centerbroken@AmS
            \gdef\broken@AmS{\vcenter{\box7}}%
          \else
             \gdef\broken@AmS{\box7}%
          \fi
      \fi
  \fi\broken@AmS}

\def\cbroken{\xdef\centerbroken@@AmS{\centerbroken@AmS}%
                       \def\centerbroken@AmS{T}\broken}               
\def\endcbroken{\endbroken\def\centerbroken@AmS{\centerbroken@@AmS}}

\def\multline#1${\in@AmS\tag{#1}\if T\cresult@AmS
 \def\multline@AmS{\def\atag@AmS{T}\getmltag@AmS#1$}\else
   \def\multline@AmS{\def\atag@AmS{F}\setbox9=\hbox{}\multline@@AmS
    \multline@@@AmS#1$}\fi\multline@AmS}                              

\def\getmltag@AmS#1\tag#2${\setbox9=\hbox{\maketag@AmS#2\tagend@AmS}%
           \multline@@AmS\multline@@@AmS#1$}

\def\multline@@AmS{\if L\tagposition@AmS
     \def\lwidth@AmS{\hskip\wd9}\def\rwidth@AmS{\hskip0pt}\else
      \def\lwidth@AmS{\hskip0pt}\def\rwidth@AmS{\hskip\wd9}\fi}      

\def\multline@@@AmS{\def\vspace##1{\noalign{\vskip ##1}}%
 \def\shoveright##1{##1\hfilneg\rwidth@AmS\quad}                      
  \def\shoveleft##1{\setbox                                           
      0=\hbox{$\displaystyle{}##1$}%
     \setbox1=\hbox{$\displaystyle##1$}%
     \ifdim\wd0=\wd1
    \hfilneg\lwidth@AmS\quad##1\else
      \setbox2=\hbox{\hskip\wd0\hskip-\wd1}%
       \hfilneg\lwidth@AmS\quad\hskip-.5\wd2 ##1\fi}
     \vbox\bgroup\Let@AmS\openup3pt\halign\bgroup\hbox to \the\displaywidth
      {$\displaystyle\hfil{}##\hfil$}\cr\hfilneg\quad
      \if L\tagposition@AmS\hskip-1em\copy9\quad\else\fi}             

\def\endmultline{\if R\tagposition@AmS\quad\box9                 
   \hskip-1em\else\fi\quad\hfilneg\crcr\egroup\egroup}

\def\aligntag#1$${\def\inaligntag@AmS{T}\openup3pt\mathsurround=0pt   
 \Let@AmS
   \def\tag{\gdef\atag@AmS{T}&}                                       
   \def\vspace##1{\noalign{\vskip##1}}                                
    \def\xtext##1{\noalign{\hbox{##1}}}                               
   \def\break{\noalign{\penalty-10000 }}                              
   \def\nobreak{\noalign{\penalty 10000 }}
   \def\allowbreak{\noalign{\penalty 0 }}
   \def\goodbreak{\noalign{\penalty -500 }}
    \gdef\atag@AmS{F}%
\if L\tagposition@AmS\laligntag@AmS#1$$\else
   \raligntag@AmS#1$$\fi}

\def\raligntag@AmS#1$${\tabskip\centering
   \halign to \the\displaywidth
{\hfil$\displaystyle{##}$\tabskip 0pt
    &$\displaystyle{{}##}$\hfil\tabskip\centering
   &\llap{\maketag@AmS##\tagend@AmS}\tabskip 0pt\cr\noalign{\vskip-
     \lineskiplimit}#1\crcr}$$}

\def\laligntag@AmS#1$${\tabskip\centering                             
   \halign to \the\displaywidth
{\hfil$\displaystyle{##}$\tabskip0pt
   &$\displaystyle{{}##}$\hfil\tabskip\centering
    &\kern-\displaywidth\rlap{\maketag@AmS##\tagend@AmS}\tabskip
    \the\displaywidth\cr\noalign{\vskip-\lineskiplimit}#1\crcr}$$}

\def\bunchtag#1$${\def\inbunchtag@AmS{T}\openup3pt\mathsurround=0pt   
    \Let@AmS
   \def\tag{\gdef\atag@AmS{T}&}
   \def\vspace##1{\noalign{\vskip##1}}
   \def\xtext##1{\noalign{\hbox{##1}}}
   \def\break{\noalign{\penalty-10000 }}
   \def\nobreak{\noalign{\penalty 10000 }}
   \def\allowbreak{\noalign{\penalty 0 }}
    \def\goodbreak{\noalign{\penalty -500 }}
  \if L\tagposition@AmS\lbunchtag@AmS#1$$\else
    \rbunchtag@AmS#1$$\fi}

\def\rbunchtag@AmS#1$${\tabskip\centering
    \halign to \displaywidth {$\hfil\displaystyle{##}\hfil$&
      \llap{\maketag@AmS##\tagend@AmS}\tabskip 0pt\cr\noalign{\vskip-
       \lineskiplimit}#1\crcr}$$}

\def\lbunchtag@AmS#1$${\tabskip\centering
   \halign to \displaywidth
{$\hfil\displaystyle{##}\hfil$&\kern-
    \displaywidth\rlap{\maketag@AmS##\tagend@AmS}\tabskip\the\displaywidth\cr
    \noalign{\vskip-\lineskiplimit}#1\crcr}$$}




\def\numeratorleft#1{#1\hskip 0pt plus 1filll\relax@AmS}
\def\numeratorright#1{\hskip 0pt plus 1filll\relax@AmS#1}
\def\numeratorcenter#1{\hskip 0pt plus 1filll\relax@AmS
      #1\hskip 0pt plus 1filll\relax@AmS}

\def\cfrac@AmS#1,{\def\numerator@AmS{#1}\cfrac@@AmS*}                 

\def\cfrac@@AmS#1;#2#3\cfend@AmS{\comp@AmS\cfmark@AmS{#2}\if T\cresult@AmS
 \gdef\cfrac@@@AmS
  {\expandafter\eat@AmS\numerator@AmS\strut\over\eat@AmS#1}\else
  \comp@AmS;{#2}\if T\cresult@AmS\gdef\cfrac@@@AmS
  {\expandafter\eat@AmS\numerator@AmS\strut\over\eat@AmS#1}\else
\gdef\cfrac@@@AmS{\if R\cftype@AmS\hfill\else\fi
    \expandafter\eat@AmS\numerator@AmS\strut
    \if L\cftype@AmS\hfill\else\fi\over
       \eat@AmS#1\displaystyle {\cfrac@AmS*#2#3\cfend@AmS}}
     \fi\fi\cfrac@@@AmS}

\def\cfrac#1{\def\cftype@AmS{C}\cfrac@AmS*#1;\cfmark@AmS\cfend@AmS}

\def\cfracl#1{\def\cftype@AmS{L}\cfrac@AmS*#1;\cfmark@AmS\cfend@AmS}

\def\cfracr#1{\def\cftype@AmS{R}\cfrac@AmS*#1;\cfmark@AmS\cfend@AmS}


\def\overrightarrow{\mathpalette\overrightarrow@AmS}

\def\overrightarrow@AmS#1#2{\vbox{\halign{$##$\cr
    #1{-}\mkern-6mu\cleaders\hbox{$#1\mkern-2mu{-}\mkern-2mu$}\hfill
     \mkern-6mu{\to}\cr
     \noalign{\kern -1pt\nointerlineskip}
     \hfil#1#2\hfil\cr}}}

\def\overleftarrow{\mathpalette\overleftarrow@Ams}

\def\overleftarrow@Ams#1#2{\vbox{\halign{$##$\cr
     #1{\leftarrow}\mkern-6mu\cleaders\hbox{$#1\mkern-2mu{-}\mkern-2mu$}\hfill
      \mkern-6mu{-}\cr
     \noalign{\kern -1pt\nointerlineskip}
     \hfil#1#2\hfil\cr}}}

\def\overleftrightarrow{\mathpalette\overleftrightarrow@AmS}

\def\overleftrightarrow@AmS#1#2{\vbox{\halign{$##$\cr
     #1{\leftarrow}\mkern-6mu\cleaders\hbox{$#1\mkern-2mu{-}\mkern-2mu$}\hfill
       \mkern-6mu{\to}\cr
    \noalign{\kern -1pt\nointerlineskip}
      \hfil#1#2\hfil\cr}}}

\def\underrightarrow{\mathpalette\underrightarrow@AmS}

\def\underrightarrow@AmS#1#2{\vtop{\halign{$##$\cr
    \hfil#1#2\hfil\cr
     \noalign{\kern -1pt\nointerlineskip}
    #1{-}\mkern-6mu\cleaders\hbox{$#1\mkern-2mu{-}\mkern-2mu$}\hfill
     \mkern-6mu{\to}\cr}}}

\def\underleftarrow{\mathpalette\underleftarrow@AmS}

\def\underleftarrow@AmS#1#2{\vtop{\halign{$##$\cr
     \hfil#1#2\hfil\cr
     \noalign{\kern -1pt\nointerlineskip}
     #1{\leftarrow}\mkern-6mu\cleaders\hbox{$#1\mkern-2mu{-}\mkern-2mu$}\hfill
      \mkern-6mu{-}\cr}}}

\def\underleftrightarrow{\mathpalette\underleftrightarrow@AmS}

\def\underleftrightarrow@AmS#1#2{\vtop{\halign{$##$\cr
      \hfil#1#2\hfil\cr
    \noalign{\kern -1pt\nointerlineskip}
     #1{\leftarrow}\mkern-6mu\cleaders\hbox{$#1\mkern-2mu{-}\mkern-2mu$}\hfill
       \mkern-6mu{\to}\cr}}}


\def\dotsc{\mathinner{\ldotp\ldotp\ldotp}}
\def\dotsi{\mathinner{\cdotp\cdotp\cdotp}}
\def\dotsj{\mathinner{\ldotp\ldotp\ldotp}}
\def\dotsb{\mathinner{\cdotp\cdotp\cdotp}}

\def\binary@AmS#1{{\thinmuskip 0mu \medmuskip 1mu \thickmuskip 1mu    
      \setbox0=\hbox{$#1{}{}{}{}{}{}{}{}{}$}\setbox1=\hbox
       {${}#1{}{}{}{}{}{}{}{}{}$}\ifdim\wd1>\wd0\gdef\binary@@AmS{T}\else
       \gdef\binary@@AmS{F}\fi}}

\def\dots{\relax@AmS\ifmmode\def\dots@AmS{\mdots@AmS}\else
    \def\dots@AmS{\tdots@AmS}\fi\dots@AmS}

\def\mdots@AmS{\futurelet\tok@AmS\mdots@@AmS}

\def\mdots@@AmS{\def\thedots@AmS{\dotsj}%
  \ifx\tok@AmS\bgroup\else
  \ifx\tok@AmS\egroup\else
  \ifx\tok@AmS$\else
  \ifx\tok@AmS\\ \iffalse}\fi\else                      
  \ifx\tok@AmS&  \iffalse}\fi\else
  \ifx\tok@AmS\left\else
  \ifx\tok@AmS\right\else
  \ifx\tok@AmS,\def\thedots@AmS{\dotsc}\else
  \inttest@AmS\tok@AmS\if T\intflag@AmS\def\thedots@AmS{\dotsi}\else
  \binary@AmS\tok@AmS\if T\binary@@AmS\def\thedots@AmS{\dotsb}\else
   \def\thedots@AmS{\dotsj}\fi\fi\fi\fi\fi\fi\fi\fi\fi\fi\thedots@AmS}

\def\tdots@AmS{\unskip\ \tdots@@AmS}

\def\tdots@@AmS{\futurelet\tok@AmS\tdots@@@AmS}

\def\tdots@@@AmS{$\ldots\,
   \ifx\tok@AmS,$\else
   \ifx\tok@AmS.\,$\else
   \ifx\tok@AmS;\,$\else
   \ifx\tok@AmS:\,$\else
   \ifx\tok@AmS?\,$\else
   \ifx\tok@AmS!\,$\else
   $\ \fi\fi\fi\fi\fi\fi}


\def\leftset#1\mid#2\rightset{\hbox{$\displaystyle
\left\{\,#1\vphantom{#1#2}\;\right|\;\left.
    #2\vphantom{#1#2}\,\right\}\offspace@AmS$}}


\def\dotii#1{{\mathop{#1}\limits^{\vbox to -1.4pt{\kern-2pt
   \hbox{\tenrm..}\vss}}}}
\def\dotiii#1{{\mathop{#1}\limits^{\vbox to -1.4pt{\kern-2pt
   \hbox{\tenrm...}\vss}}}}
\def\dotiv#1{{\mathop{#1}\limits^{\vbox to -1.4pt{\kern-2pt
   \hbox{\tenrm....}\vss}}}}

\def\hatsymbol{{\mathchoice{\null}{\null}{\,\,\hbox{\lower 10pt\hbox
    {$\widehat{\null}$}}}{\,\hbox{\lower 20pt\hbox
       {$\hat{\null}$}}}}}


\def\overset#1\to#2{{\mathop{#2}^{#1}}}

\def\underset#1\to#2{{\mathop{#2}_{#1}}}

\def\oversetbrace#1\to#2{{\overbrace{#2}^{#1}}}
\def\undersetbrace#1\to#2{{\underbrace{#2}_{#1}}}


\def\theuproot{0 pt}

\def\therightroot{0mu}

\def\r@@t#1#2{\setbox\z@\hbox{$\m@th#1\sqrt{#2}$}%
  \dimen@\ht\z@ \advance\dimen@-\dp\z@ \advance\dimen@\theuproot
  \mskip5mu\raise.6\dimen@\copy\rootbox \mskip-10mu \mskip\therightroot
    \box\z@\gdef\theuproot{0 pt}\gdef\therightroot{0mu}}              


\def\boxed#1{\setbox0=\hbox{$\displaystyle{#1}$}\hbox{\lower.4pt\hbox{\lower
   3pt\hbox{\lower 1\dp0\hbox{\vbox{\hrule height .4pt \hbox{\vrule width
   .4pt \hskip 3pt\vbox{\vskip 3pt\box0\vskip3pt}\hskip 3pt \vrule width
      .4pt}\hrule height .4pt}}}}}}


\def\documentstyle#1{\input #1.sty}

\newif\ifretry@AmS
\def\y@AmS{y } \def\y@@AmS{Y } \def\n@AmS{n } \def\n@@AmS{N }
\def\ask@AmS{\message
  {Do you want output? (y or n, follow answer by return) }\loop
   \read-1 to\answer@AmS
  \ifx\answer@AmS\y@AmS\retry@AmSfalse\outputon
   \else\ifx\answer@AmS\y@@AmS\retry@AmSfalse\outputon
    \else\ifx\answer@AmS\n@AmS\retry@AmSfalse\outputoff
     \else\ifx\answer@AmS\n@@AmS\retry@AmSfalse\outputoff
      \else \retry@AmStrue\fi\fi\fi\fi
  \ifretry@AmS\message{Type y or n, follow answer by return: }\repeat}

\def\outputoff{\global\output{\setbox0=\box255 \deadcycles=0}}

\def\outputon{\global\output{\output@AmS}}

\catcode`\@=13


\catcode`\@=11



\normallineskiplimit=1pt
\parindent 10pt
\hsize 26pc
\vsize 42pc


\font\eightrm=cmr8
\font\sixrm=cmr6
\font\eighti=cmmi8 \skewchar\eighti='177
\font\sixi=cmmi6 \skewchar\sixi='177
\font\eightsy=cmsy8 \skewchar\eightsy='60
\font\sixsy=cmsy6 \skewchar\sixsy='60
\font\eightbf=cmbx8
\font\sixbf=cmbx6
\font\eightsl=cmsl8
\font\eightit=cmti8
\font\tensmc=cmcsc10


\font\ninerm=cmr9
\font\ninei=cmmi9 \skewchar\ninei='177
\font\ninesy=cmsy9 \skewchar\ninesy='60
\font\ninebf=cmbx9
\font\ninesl=cmsl9
\font\nineit=cmti9


\def\tenpoint{\def\pointsize@AmS{t}\normalbaselineskip=12pt            
 \abovedisplayskip 12pt plus 3pt minus 9pt
 \belowdisplayskip 12pt plus 3pt minus 9pt
 \abovedisplayshortskip 0pt plus 3pt
 \belowdisplayshortskip 7pt plus 3pt minus 4pt
 \def\rm{\fam0\tenrm}%
 \def\it{\fam\itfam\tenit}%
 \def\sl{\fam\slfam\tensl}%
 \def\bf{\fam\bffam\tenbf}%
 \def\smc{\tensmc}%
 \def\mit{\fam 1}%
 \def\cal{\fam 2}%
 \textfont0=\tenrm   \scriptfont0=\sevenrm   \scriptscriptfont0=\fiverm
 \textfont1=\teni    \scriptfont1=\seveni    \scriptscriptfont1=\fivei
 \textfont2=\tensy   \scriptfont2=\sevensy   \scriptscriptfont2=\fivesy
 \textfont3=\tenex   \scriptfont3=\tenex     \scriptscriptfont3=\tenex
 \textfont\itfam=\tenit
 \textfont\slfam=\tensl
 \textfont\bffam=\tenbf \scriptfont\bffam=\sevenbf
   \scriptscriptfont\bffam=\fivebf
\normalbaselines\rm}

\def\eightpoint{\def\pointsize@AmS{8}\normalbaselineskip=10pt
 \abovedisplayskip 10pt plus 2.4pt minus 7.2pt
 \belowdisplayskip 10pt plus 2.4pt minus 7.2pt
 \abovedisplayshortskip 0pt plus 2.4pt
 \belowdisplayshortskip 5.6pt plus 2.4pt minus 3.2pt
 \def\rm{\fam0\eightrm}%
 \def\it{\fam\itfam\eightit}%
 \def\sl{\fam\slfam\eightsl}%
 \def\bf{\fam\bffam\eightbf}%
 \def\mit{\fam 1}%
 \def\cal{\fam 2}%
 \textfont0=\eightrm   \scriptfont0=\sixrm   \scriptscriptfont0=\fiverm
 \textfont1=\eighti    \scriptfont1=\sixi    \scriptscriptfont1=\fivei
 \textfont2=\eightsy   \scriptfont2=\sixsy   \scriptscriptfont2=\fivesy
 \textfont3=\tenex   \scriptfont3=\tenex     \scriptscriptfont3=\tenex
 \textfont\itfam=\eightit
 \textfont\slfam=\eightsl
 \textfont\bffam=\eightbf \scriptfont\bffam=\sixbf
   \scriptscriptfont\bffam=\fivebf
\normalbaselines\rm}


\def\ninepoint{\def\pointsize@AmS{9}\normalbaselineskip=11pt
 \abovedisplayskip 11pt plus 2.7pt minus 8.1pt
 \belowdisplayskip 11pt plus 2.7pt minus 8.1pt
 \abovedisplayshortskip 0pt plus 2.7pt
 \belowdisplayshortskip 6.3pt plus 2.7pt minus 3.6pt
 \def\rm{\fam0\ninerm}%
 \def\it{\fam\itfam\nineit}%
 \def\sl{\fam\slfam\ninesl}%
 \def\bf{\fam\bffam\ninebf}%
 \def\mit{\fam 1}%
 \def\cal{\fam 2}%
 \textfont0=\ninerm   \scriptfont0=\sevenrm   \scriptscriptfont0=\fiverm
 \textfont1=\ninei    \scriptfont1=\seveni    \scriptscriptfont1=\fivei
 \textfont2=\ninesy   \scriptfont2=\sevensy   \scriptscriptfont2=\fivesy
 \textfont3=\tenex   \scriptfont3=\tenex     \scriptscriptfont3=\tenex
 \textfont\itfam=\nineit
 \textfont\slfam=\ninesl
 \textfont\bffam=\ninebf \scriptfont\bffam=\sevenbf
   \scriptscriptfont\bffam=\fivebf
\normalbaselines\rm}


\newcount\footmarkcount@AmS
\footmarkcount@AmS=0
\newcount\foottextcount@AmS
\foottextcount@AmS=0

\def\footnotemark{\unskip\futurelet\tok@AmS\footnotemark@AmS}
\def\footnotemark@AmS{\ifx [\tok@AmS \def\next@AmS{\footnotemark@@AmS}\else
   \def\next@AmS{\footnotemark@@@AmS}\fi\next@AmS}
\def\footnotemark@@AmS[#1]{{#1}}
\def\footnotemark@@@AmS{\global\advance\footmarkcount@AmS by 1
 \xdef\thefootmarkcount@AmS{\the\footmarkcount@AmS}$^{\thefootmarkcount@AmS}$}

\def\makefootnote@AmS#1#2{\insert\footins{\interlinepenalty100
   \eightpoint
  \splittopskip=6.8pt
  \splitmaxdepth=2.8pt
   \floatingpenalty=20000
   \leftskip = 0pt  \rightskip = 0pt
    \noindent {#1}\footstrut{\ignorespaces#2\unskip}\topsmash{\strut}}}

\def\footnotetext{\futurelet\tok@AmS\footnotetext@}
\def\footnotetext@{\ifx [\tok@AmS \def\next@AmS{\footnotetext@@AmS}\else
  \def\next@AmS{\footnotetext@@@AmS}\fi\next@AmS}
\def\footnotetext@@AmS[#1]#2{\makefootnote@AmS{#1}{#2}}
\def\footnotetext@@@AmS#1{\global\advance\foottextcount@AmS by 1
  \xdef\thefoottextcount@AmS{\the\foottextcount@AmS}%
\makefootnote@AmS{$^{\thefoottextcount@AmS}$}{#1}}

\def\footnote{\unskip\futurelet\tok@AmS\footnote@AmS}
\def\footnote@AmS{\ifx [\tok@AmS \def\next@AmS{\footnote@@AmS}\else
   \def\next@AmS{\footnote@@@AmS}\fi\next@AmS}
\def\footnote@@AmS[#1]#2{{\edef\sf{\the\spacefactor}%
  {#1}\makefootnote@AmS{#1}{#2}\spacefactor=\sf}}
\def\footnote@@@AmS#1{\ifnum\footmarkcount@AmS=\foottextcount@AmS\else
 \errmessage{AmS-TeX warning: last footnote marker was \the\footmarkcount@AmS,
   last footnote was
   \the\foottextcount@AmS}\footmarkcount@AmS=\foottextcount@AmS\fi
   {\edef\sf{\the\spacefactor}\footnotemark@@@AmS\footnotetext@@@AmS{#1}%
    \spacefactor=\sf}}

\def\adjustfootnotemark#1{\advance\footmarkcount@AmS by #1}           
\def\adjustfootnote#1{\advance\foottextcount@AmS by #1}


\def\topmatter@AmS{F}                                                 
\def\topmatter{\def\topmatter@AmS{T}}

\def\filhss@AmS{plus 1000pt}                                          
\def\overlong{\def\filhss@AmS{plus 1000pt minus1000pt}}

\newbox\titlebox@AmS

\setbox\titlebox@AmS=\vbox{}                                          

\def\title#1\endtitle{{\let\\=\cr                                     
  \global\setbox\titlebox@AmS=\vbox{\tabskip0pt\filhss@AmS
  \halign to \hsize
    {\tenpoint\bf\hfil\ignorespaces##\unskip\hfil\cr#1\cr}}}\def     
     \filhss@AmSs{plus 1000pt}}

\def\isauthor@AmS{F}                                                
\newbox\authorbox@AmS

\def\author#1\endauthor{\gdef\isauthor@AmS{T}{\let\\=\cr
 \global\setbox\authorbox@AmS=\vbox{\tabskip0pt
 \filhss@AmS\halign to \hsize
   {\tenpoint\smc\hfil\ignorespaces##\unskip\hfil\cr#1\cr}}}\def
      \filhss@AmS{plus 1000pt}}


\def\uctext@AmS#1{\uppercase@AmS#1\gdef                           
       \uppercase@@AmS{}${\hskip-2\mathsurround}$}
\def\uppercase@AmS#1$#2${\gdef\uppercase@@AmS{\uppercase@AmS}\uppercase
    {#1}${#2}$\uppercase@@AmS}

\newcount\Notes@AmS                                             

\def\sfootnote@AmS{\unskip\futurelet\tok@AmS\sfootnote@@AmS}
\def\sfootnote@@AmS{\ifx [\tok@AmS \def\next@AmS{\sfootnote@@@AmS}\else
    \def\next@AmS{\sfootnote@@@@AmS}\fi\next@AmS}
\def\sfootnote@@@AmS[#1]#2{\global\toks@{#2}\advance\Notes@AmS by 1
  \expandafter\xdef\csname Note\romannumeral\Notes@AmS @AmS\endcsname
   {\the\toks@}}
\def\sfootnote@@@@AmS#1{\global\toks@{#1}\global\advance\Notes@AmS by 1
  \expandafter\xdef\csname Note\romannumeral\Notes@AmS @AmS\endcsname
  {\the\toks@}}

\def\Sfootnote@AmS{\unskip\futurelet\tok@AmS\Sfootnote@@AmS}
\def\Sfootnote@@AmS{\ifx [\tok@AmS \def\next@AmS{\Sfootnote@@@AmS}\else
    \def\next@AmS{\Sfootnote@@@@AmS}\fi\next@AmS}
\def\Sfootnote@@@AmS[#1]#2{{#1}\advance\Notes@AmS by 1
  {\edef\sf{\the\spacefactor}\makefootnote@AmS{#1}{\csname
     Note\romannumeral\Notes@AmS @AmS\endcsname}\spacefactor=\sf}}
\def\Sfootnote@@@@AmS#1{\ifnum\footmarkcount@AmS=\foottextcount@AmS\else
 \errmessage{AmS-TeX warning: last footnote marker was \the\footmarkcount@AmS,
  last footnote was
   \the\foottextcount@AmS}\footmarkcount@AmS=\foottextcount@AmS\fi
 {\edef\sf{\the\spacefactor}\footnotemark@@@AmS \global\advance\Notes@AmS by 1
    \footnotetext@@@AmS{\csname
      Note\romannumeral\Notes@AmS @AmS\endcsname}\spacefactor=\sf}}

\def\TITLE#1\endTITLE                                           
{{\Notes@AmS=0 \let\\=\cr\let\footnote=\sfootnote@AmS
   \setbox0=\vbox{\tabskip\centering
  \halign to \hsize{\tenpoint\bf\ignorespaces##\unskip\cr#1\cr}}
 \Notes@AmS=0   \let\footnote=\Sfootnote@AmS
   \global\setbox\titlebox@AmS=\vbox{\tabskip0pt\filhss@AmS
\halign to \hsize{\tenpoint\bf\hfil
 \uctext@AmS{\ignorespaces##\unskip}\hfil\cr
          #1\cr}}}\def\filhss@AmS{plus 1000pt}}

\def\AUTHOR#1\endAUTHOR{\gdef\isauthor@AmS{T}{\Notes@AmS=0 \let\\=\cr
   \let\footnote=\sfootnote@AmS
 \setbox0 =\vbox{\tabskip\centering\halign to \hsize{\tenpoint\smc
   \ignorespaces##\unskip\cr#1\cr}}\Notes@AmS=0
   \let\footnote=\Sfootnote@AmS
  \global\setbox\authorbox@AmS=\vbox{\tabskip0pt\filhss@AmS\halign
  to \hsize{\tenpoint\smc\hfil\uppercase{\ignorespaces
     ##\unskip}\hfil\cr#1\cr}}}\def\filhss@AmS{plus 1000pt}}


\newcount\language@AmS                                            
\language@AmS=0
\def\german{\language@AmS=1}

\def\abstractword@AmS{\ifcase \language@AmS ABSTRACT\or ZUSAMMENFASSUNG\fi}
\def\logoword@AmS{\ifcase \language@AmS Typeset by \fi}
\def\subjclassword@AmS{\ifcase \language@AmS
     1980 Mathematics subject classifications \fi}
\def\keywordsword@AmS{\ifcase \language@AmS  Keywords and phrases\fi}
\def\Referenceword@AmS{\ifcase \language@AmS References\fi}

\def\isaffil@AmS{F}
\newbox\affilbox@AmS
\def\affil{\gdef\isaffil@AmS{T}\bgroup\let\\=\cr
   \global\setbox\affilbox@AmS
     =\vbox\bgroup\tabskip0pt\filhss@AmS
 \halign to \hsize\bgroup\tenpoint\hfil\ignorespaces##\unskip\hfil\cr}

\def\endaffil{\cr\egroup\egroup\egroup\def\filhss@AmS{plus 1000pt}}

\newcount\addresscount@AmS                                         
\addresscount@AmS=0

\def\address#1{\global\advance\addresscount@AmS by 1
  \expandafter\gdef\csname address\romannumeral\addresscount@AmS\endcsname
   {\noindent\eightpoint\ignorespaces#1\par}}

\def\isdate@AmS{F}                                                 
\def\date#1{\gdef\isdate@AmS{T}\gdef\date@AmS{\tenpoint\ignorespaces#1\unskip}}

\def\isthanks@AmS{F}
\def\thanks#1{\gdef\isthanks@AmS{T}\gdef\thanks@AmS{\eightpoint\ignorespaces
       #1\unskip}}

\def\keywords@AmS{}                                                
\def\keywords#1{\def\keywords@AmS{\noindent \eightpoint \it
\keywordsword@AmS .\enspace \rm\ignorespaces#1\par}}

\def\subjclass@AmS{}
\def\subjclass#1{\def\subjclass@AmS{\noindent \eightpoint\it
\subjclassword@AmS
(Amer.\ Math.\ Soc.)\/\rm: \ignorespaces#1\par}}

\def\isabstract@AmS{F}
\long\def\abstract#1{\gdef\isabstract@AmS{T}\long\gdef\abstract@AmS
   {\eightpoint \abstractword@AmS\period\ignorespaces #1\par}}        


\def\pretitle{}
\def\preauthor{}
\def\preaffil{}
\def\predate{}
\def\preabstract{}
\def\prepaper{}


\def\endtopmatter{\if F\topmatter@AmS \errmessage{AmS-TeX warning: You
    forgot the \string\topmatter, but I forgive you.}\fi
\hrule height 0pt \vskip -\topskip                                   
   \pretitle
   \vskip 24pt plus 12pt minus 12pt
   \unvbox\titlebox@AmS                                              
   \preauthor
   \if T\isauthor@AmS \vskip 12pt plus 6pt minus 3pt
       \unvbox\authorbox@AmS \else\fi
    \preaffil
   \if T\isaffil@AmS \vskip 10pt plus 5pt minus 2pt
       \unvbox\affilbox@AmS\else\fi
  \predate
   \if T\isdate@AmS \vskip 6pt plus 2pt minus 1pt
  \hbox to \hsize{\hfil\date@AmS\hfil}\else\fi
    \preabstract
\if T\isthanks@AmS
  \makefootnote@AmS{}{\thanks@AmS}\else\fi
   \if T\isabstract@AmS \vskip 15pt plus 12pt minus 12pt
 {\leftskip=16pt\rightskip=16pt
  \noindent \abstract@AmS}\else\fi
   \prepaper
     \vskip 18pt plus 12pt minus 6pt \tenpoint}


\newcount\addresnum@AmS                                               
\def\enddocument{\penalty10000 \sfcode`\.3000\vskip 12pt minus 6pt  
\keywords@AmS                                                         
\subjclass@AmS
\addresnum@AmS=0
  \loop\ifnum\addresnum@AmS<\addresscount@AmS\advance\addresnum@AmS by 1
  \csname address\romannumeral\addresnum@AmS\endcsname\repeat
\vfill\supereject\end}


\newbox\headingbox@AmS
\outer\def\heading{\medbreak\bgroup\let\\=\cr
\global\setbox\headingbox@AmS=\vbox\bgroup\tabskip0pt\filhss@AmS      
   \halign to \hsize\bgroup\tenpoint\smc\hfil\ignorespaces
            ##\unskip\hfil\cr}

\def\endheading{\cr\egroup\egroup\egroup\unvbox\headingbox@AmS
    \penalty10000 \def\filhss@AmS{plus 1000pt}\medskip}


\outer\def\proclaim#1{\xdef\curfont@AmS{\the\font}\medbreak        
  \noindent\smc\ignorespaces#1\unskip.\enspace\sl\ignorespaces}

\outer\def\proclaimnp#1{\xdef\curfont@AmS{\the\font}\medbreak      
  \noindent\smc\ignorespaces#1\enspace\sl\ignorespaces}

\def\finishproclaim{\par\curfont@AmS\ifdim\lastskip<\medskipamount 
 \removelastskip \penalty 55\medskip\fi}

\outer\def\demo#1{\par\ifdim\lastskip<\smallskipamount
  \removelastskip\smallskip\fi\noindent{\smc\ignorespaces#1\unskip:}\enspace
     \ignorespaces}

\outer\def\demonp#1{\ifdim\lastskip<\smallskipamount
  \removelastskip\smallskip\fi\noindent{\smc#1}\enspace\ignorespaces}

\newif\ifrunin@AmS                                                    
\runin@AmSfalse
\def\runin{\runin@AmStrue}
\def\conditions{\def\\##1:{\par\noindent                              
   \hbox to 1.5\parindent{\hss\rm\ignorespaces##1\unskip}%
      \hskip .5\parindent \hangafter1\hangindent2\parindent\ignorespaces}%
    \def\firstcon@AmS##1:{\ifrunin@AmS
     {\rm\ignorespaces##1\unskip}\ \ignorespaces
  \else\par\ifdim\lastskip<\smallskipamount\removelastskip\penalty55
     \smallskip\fi
     \\##1:\fi}\firstcon@AmS}
\def\endconditions{\par\smallbreak\runin@AmSfalse}                    


\def\refto#1{\in@AmS,{#1}\if T\cresult@AmS\refto@AmS#1\end@AmS\else   
    [{\bf#1}]\fi}
\def\refto@AmS#1,#2\end@AmS{[{\bf#1},#2]}

\def\Refs{\bigbreak\hbox to \hsize{\hfil\tenpoint
    \smc \Referenceword@AmS\hfil}\penalty 10000
      \bigskip\eightpoint\sfcode`.=1000 }                             

\newbox\nobox@AmS        \newbox\keybox@AmS        \newbox\bybox@AmS  
\newbox\bysamebox@AmS    \newbox\paperbox@AmS      \newbox\paperinfobox@AmS
\newbox\jourbox@AmS      \newbox\volbox@AmS        \newbox\issuebox@AmS
\newbox\yrbox@AmS        \newbox\pagesbox@AmS      \newbox\bookbox@AmS
\newbox\bookinfobox@AmS  \newbox\publbox@AmS       \newbox\publaddrbox@AmS
\newbox\finalinfobox@AmS

\def\refset@AmS#1{\expandafter\gdef\csname is\expandafter\eat@AmS     
  \string#1@AmS\endcsname{F}\expandafter
  \setbox\csname \expandafter\eat@AmS\string#1box@AmS\endcsname=\null}

\def\ref@AmS{\refset@AmS\no \refset@AmS\key \refset@AmS\by            
\gdef\isbysame@AmS{F}
 \refset@AmS\paper
  \refset@AmS\paperinfo \refset@AmS\jour \refset@AmS\vol
  \refset@AmS\issue \refset@AmS\yr
  \gdef\istoappear@AmS{F}
  \refset@AmS\pages
  \gdef\ispage@AmS{F}
  \refset@AmS\book
  \gdef\isinbook@AmS{F}
  \refset@AmS\bookinfo \refset@AmS\publ
  \refset@AmS\publaddr \refset@AmS\finalinfo \bgroup
     \ignorespaces}                                                   

\def\ref{\noindent\hangindent 20pt \hangafter 1 \def\refi@AmS{T}
  \def\refl@AmS{F}\def\\{\egroup\endref@AmS\gdef\refi@AmS{F}\ref@AmS}\ref@AmS}

\def\refdef@AmS#1#2{\def#1{\egroup\expandafter                        
  \gdef\csname is\expandafter\eat@AmS
  \string#1@AmS\endcsname{T}\expandafter\setbox
   \csname \expandafter\eat@AmS\string#1box@AmS\endcsname=\hbox\bgroup#2}}

\refdef@AmS\no{} \refdef@AmS\key{} \refdef@AmS\by{}
\def\bysame{\egroup\gdef\isbysame@AmS{T}\bgroup}                    
\refdef@AmS\paper\it
\refdef@AmS\paperinfo{} \refdef@AmS\jour{} \refdef@AmS\vol\bf
\refdef@AmS\issue{} \refdef@AmS\yr{}
\def\toappear{\egroup\gdef\istoappear@AmS{T}\bgroup}                
\refdef@AmS\pages{}
\def\page{\egroup\gdef\ispage@AmS{T}\setbox
                 \pagesbox@AmS=\hbox\bgroup}                        
\refdef@AmS\book{}
\def\inbook{\egroup\gdef\isinbook@AmS{T}\setbox
                               \bookbox@AmS=\hbox\bgroup}           
\refdef@AmS\bookinfo{} \refdef@AmS\publ{}
\refdef@AmS\publaddr{}
\refdef@AmS\finalinfo{}

\def\setpunct@AmS{\def\prepunct@AmS{, }}                              
\def\ppunbox@AmS#1{\prepunct@AmS\unhbox#1\unskip}                     

\def\endref@AmS{\def\prepunct@AmS{}
\if T\refi@AmS                                                      
  \if F\isno@AmS\hbox to 10pt{}\else                                
     \hbox to 20pt{\hss\unhbox\nobox@AmS\unskip. }\fi               
  \if T\iskey@AmS \unhbox\keybox@AmS\unskip\ \fi                    
  \if T\isby@AmS  \hbox{\unhcopy\bybox@AmS\unskip}\setpunct@AmS     
         \setbox\bysamebox@AmS=\hbox{\unhcopy\bybox@AmS\unskip}\fi  
  \if T\isbysame@AmS                                                
   \hbox to \wd\bysamebox@AmS{\leaders\hrule\hfill}\setpunct@AmS\fi
 \fi                                                                
  \if T\ispaper@AmS\ppunbox@AmS\paperbox@AmS\setpunct@AmS\fi          
  \if T\ispaperinfo@AmS\ppunbox@AmS\paperinfobox@AmS\setpunct@AmS\fi  
  \if T\isjour@AmS\ppunbox@AmS\jourbox@AmS\setpunct@AmS               
     \if T\isvol@AmS \ \unhbox\volbox@AmS\unskip\setpunct@AmS\fi    
     \if T\isissue@AmS \ \unhbox\issuebox@AmS\unskip\setpunct@AmS\fi
     \if T\isyr@AmS \ (\unhbox\yrbox@AmS\unskip)\setpunct@AmS\fi    
     \if T\istoappear@AmS \ (to appear)\setpunct@AmS\fi             
     \if T\ispages@AmS \ppunbox@AmS\pagesbox@AmS\setpunct@AmS\fi    
     \if T\ispage@AmS                                               
           \prepunct@AmS p.\ \unhbox\pagesbox@AmS\unskip\setpunct@AmS\fi
     \fi                                                            
  \if T\isbook@AmS \prepunct@AmS                                      
                     ``\unhbox\bookbox@AmS\unskip''\setpunct@AmS\fi
  \if T\isinbook@AmS \prepunct@AmS                                    
    \unskip\ in ``\unhbox\bookbox@AmS\unskip''\setpunct@AmS
       \gdef\isbook@AmS{T}\fi
  \if T\isbookinfo@AmS \ppunbox@AmS\bookinfobox@AmS\setpunct@AmS\fi   
  \if T\ispubl@AmS \ppunbox@AmS\publbox@AmS\setpunct@AmS\fi           
  \if T\ispubladdr@AmS \ppunbox@AmS\publaddrbox@AmS\setpunct@AmS\fi   
 \if T\isbook@AmS                                                     
  \if T\isyr@AmS \prepunct@AmS \unhbox\yrbox@AmS\unskip             
              \setpunct@AmS\fi
  \if T\istoappear@AmS \ (to appear)\setpunct@AmS\fi                
  \if T\ispages@AmS                                                 
    \prepunct@AmS pp.\ \unhbox\pagesbox@AmS\unskip\setpunct@AmS\fi
  \if T\ispage@AmS                                                  
    \prepunct@AmS p.\ \unhbox\pagesbox@AmS\unskip\setpunct@AmS\fi
 \fi
  \if T\isfinalinfo@AmS \period\unhbox\finalinfobox@AmS\else          
    \if T\refl@AmS .\else ; \fi\fi}

\def\endref{\egroup\gdef\refl@AmS{T}\endref@AmS\par}


\newif\ifguides@AmS
\guides@AmSfalse
\def\guidelines{\guides@AmStrue}
\def\noguidelines{\guides@AmSfalse}
\def\guidelinegap#1{\def\gwidth@AmS{#1}}
\def\gwidth@AmS{24pt}

\newif\iflogo@AmS
\def\nologo{\logo@AmSfalse}
\logo@AmStrue

\def\output@AmS{\ifnum\count0=1
 \shipout\vbox{\ifguides@AmS\hrule width \hsize \vskip\gwidth@AmS \fi
   \vbox to \vsize{\boxmaxdepth=\maxdepth\pagecontents}\baselineskip2pc
\iflogo@AmS \hbox to \hsize{\hfil\eightpoint \logoword@AmS\AmSTeX}\fi
     \ifguides@AmS \vskip\gwidth@AmS
\hrule width \hsize\fi}\vsize 44pc\else
   \shipout\vbox{\ifguides@AmS \hrule width \hsize \vskip\gwidth@AmS\fi
   \vbox to \vsize{\boxmaxdepth=\maxdepth\pagecontents}\baselineskip2pc\hbox to
  \hsize{\hfil \tenpoint\number\count0\hfil}\ifguides@AmS
    \vskip\gwidth@AmS\hrule width \hsize\fi}\fi\global\advance\count0 by 1
  \global\footmarkcount@AmS=0 \global\foottextcount@AmS=0
 \ifnum\outputpenalty>-20000 \else\dosupereject\fi}




\tenpoint

\catcode`\@=13

\output={\plainoutput}

\magnification=\magstep1
\baselineskip=16pt
\hoffset=-0.75truecm
\voffset=0.0truecm
\vsize=23.5truecm
\hsize=18.0truecm
\parskip=0.2cm
\parindent=1cm

\hfuzz=23pt

\def \bigbreak  {\goodbreak\bigskip}
\def \medbreak  {\goodbreak\medskip}
\def \smallbreak{\goodbreak\smallskip}
\def \header#1{\goodbreak\bigskip\centerline{\bf #1}\medskip\nobreak}
\def \subheader#1{\goodbreak\medskip\par\noindent{\bf #1}\smallskip\nobreak}

%
%
\def\pmb#1{\setbox0=\hbox{#1}%
  \kern-.025em\copy0\kern-\wd0
  \kern.05em\copy0\kern-\wd0
  \kern-.025em\raise.0433em\box0 }
%
%
%
\def\timedate{ {\tt
\count215=\time \divide\count215 by60  \number\count215
\multiply\count215 by-60 \advance \count215 by\time :\number\count215 \space
\number\day\space
\ifcase\month\or January\or February\or March\or April\or May\or June\or July
\or August\or September\or October\or November\or December\fi\space\number\year
}}
%

%
%
\def\captpar{\dimen0=\hsize
             \advance\dimen0 by -1.0truecm
             \par\parshape 1 0.5truecm \dimen0 \noindent}
\def\pp{\dimen0=\hsize
        \advance\dimen0 by -1truecm
        \par\parshape 2 0truecm \dimen0 1truecm \dimen0 \noindent}
%
%
\def\apjpap#1;#2;#3;#4; {\pp#1, {\sl #2}, {\bf #3}, #4.}
%
%
\def\apjbook#1;#2;#3;#4; {\pp#1, {\sl #2} (#3: #4).}
%
%
\def\apjppt#1;#2; {\pp#1, #2.}
\def\s {\scriptscriptstyle}
\def\ccdot{{\hskip-0.7pt\cdot\hskip-0.7pt}}

\def\sqr#1#2{{\vcenter{\hrule height.#2pt
              \hbox{\vrule width.#2pt height#1pt \kern#1pt \vrule width.#2pt}
              \hrule height.#2pt}}}

\def\mathrelfun#1#2{\lower3.6pt\vbox{\baselineskip0pt\lineskip.9pt
  \ialign{$\mathsurround=0pt#1\hfil##\hfil$\crcr#2\crcr\sim\crcr}}}
\def\simlt{\mathrel{\mathpalette\mathrelfun <}}
\def\simgt{\mathrel{\mathpalette\mathrelfun >}}

\def\ln {{\rm ln}}
\def\sgn {{\rm sgn}}

\def\erf {{\rm erf}}

\def\rmc {{\rm c}}
\def\rmd {{\rm d}}

\def\rms {{\rm s}}

\def\rmx {{\rm x}}
\def\rmy {{\rm y}}

\def\rmA {{\rm A}}

\def\rmP {{\rm P}}

\def\bfk {{\bf k}}

\def\bfn {{\bf n}}

\def\bfx {{\bf x}}

\def\calP {{\cal P}}

\def\hatbfn  {{\hat\bfn}}

\def\km  {{\rm km}}

\def\Mpc {{\rm Mpc}}

\def\eV  {{\rm \hbox{e\kern-0.14em V}}}
\def\keV {{\rm \hbox{ke\kern-0.14em V}}}
\def\MeV {{\rm \hbox{Me\kern-0.14em V}}}
\def\GeV {{\rm \hbox{Ge\kern-0.14em V}}}

\def\Rcurv{{R_{\rm curv}}}
\def\etaobs{{\eta_{\rm obs}}}
\def\etals {{\eta_{\rm ls}}}

\def\corrfuncSI#1{{\Xi_{{}_{\scriptstyle{#1}}}}}

\def\corrfunc#1{{\xi_{{}_{\scriptstyle{#1}}}}}

\def\Robs{{R_{\rm obs}}}

\input psfig

\font\FermiPPTfont=cmssbx10 scaled 1440
\font\FermiSmallfont=cmssq8 scaled 1200

\def\FNALpptheadnologo#1#2{
\null \vskip -1truein
\centerline{\hbox to 7.5truein {
\hskip 1.5cm
\vbox to 1in{\vfill
             \hbox{{\FermiPPTfont Fermi National Accelerator Laboratory}}
             \vfill}
\hfill
\vbox to 1in {\vfill \FermiSmallfont
              \hbox{#1}
              \hbox{#2}
              \vfill}
}}}

\FNALpptheadnologo{FERMILAB-Pub-94/386-A}{December 1994}

\topmatter
\title
No Very Large Scale Structure in an Open Universe
\footnote{\tt DAMTP preprint R94/46,
(Submitted to {\smc The Physical Review D}) }
\endtitle
\author
{Albert Stebbins$^\spadesuit$} {\rm and}
{R.R. Caldwell$^{\spadesuit\clubsuit}$}
\endauthor
\affil
$\spadesuit$ NASA/Fermilab Astrophysics Center,
FNAL, Box 500, Batavia, Illinois 60510, USA \\
$\clubsuit$ University of Cambridge, D.A.M.T.P.
Silver Street, Cambridge, England, CB3 9EW
\endaffil
\abstract{\ninepoint We study the effects of negative spatial curvature on the
statistics of
inhomogeneities in open cosmological models.  In particular we examine the
suppression of large-separation correlations in density and gravitational
potential fluctuations and the resulting suppression of large-angle
correlations in the anisotropy of the microwave background radiation.  We
obtain an expression which gives the {\it minimum} amount of suppression of
correlations for any statistical distribution described by a ``power
spectrum''. This minimum suppression requires that the correlations fall off
exponentially above the curvature scale.  To the extent that the observed
correlations in the temperature anisotropy extend to large angular scales,
one can set a lower bound to the radius of curvature and hence on $\Omega_0$.
}
\endtopmatter

\header{1. Introduction}

	In this paper we examine the effects of negative spatial curvature on
the statistics of inhomogeneities and in particular the microwave background
radiation (MBR) anisotropy.  Throughout this work, we shall consider only a
cosmology described by a homogeneous, isotropic Friedmann-Robertson-Walker
(FRW) spacetime with negative spatial curvature. The spatial geometry is that
of $H^3$, a 3-hyperboloid.  We may call such an expanding spacetime ``open''
because the spatial manifold, $H^3$, is non-compact.

	In an open FRW spacetime the spatial sections are intrinsically
curved with a fixed curvature radius $R_{\rm curve}$ at all points in space at
a given time, such that the spatial Ricci scalar is
${}^3{\rm R} = 6 / R^2_{\rm curve}$. On length scales much smaller than the
radius of curvature,
$l\ll R_{\rm curve}$, the space effectively has a flat, Euclidean geometry.
On length scales $l\simgt R_{\rm curve}$, the space has a hyperbolic,
Lobachevskiian geometry. Hence, the effects of spatial curvature ought to
be manifest in physical processes on length scales comparable to the radius
of curvature.

         Neither classical tests for curvature such as number counts and the
redshift-distance relation, nor more modern techniques such as the statistics
of gravitational lenses or anisotropy of clustering in redshift space, have yet
to provide any conclusive evidence for or against the presence of spatial
curvature [1].  Each of these tests attempts to probe the geometry directly, or
indirectly through the effect of spatial curvature on the expansion of the
universe.  Here we examine a different consequence of the curvature, namely how
the curvature influences the correlations of the inhomogeneities.

	In particular we shall show that for any spectrum of inhomogeneities
that correlations of the inhomogeneities must fall off exponentially for
spatial separations greater than the radius of curvature in an open cosmology
(see ref~[2] for another discussion of these effects).  Such a fall-off is also
possible in a flat or closed cosmology, subject to the specific behavior of the
spectrum of fluctuations. Thus one cannot determine the curvature scale without
making further assumptions about the spectrum of inhomogeneities [3,4,5].
Nevertheless, we may set a lower bound on the radius of curvature through this
technique.

	The plan of the paper is as follows.  In \S2 we discuss the
relationship between correlation lengths of functions and the wavenumbers
which parameterize the integral transforms of the functions.  In \S3 we show
that for open cosmological models in which the inhomogeneities are described by
a power spectrum there is necessarily, {\it on average}, a suppression of
very-large-scale correlations in the inhomogeneity. Here ``very-large'' means
larger than the curvature scale. It is shown in \S4 how the suppression of
large-scale power may be used to set a lower limit on the curvature scale from
observations of MBR anisotropies.  In \S5 we examine a toy model of
inhomogeneities with very-large-scale spatial structure but small-scale angular
structure in the observed MBR anisotropies.

        While open cosmologies have been studied by many authors the basic
mathematical formulae used to describe functions in open cosmologies are
probably unfamiliar to many readers.  We neither wish to inundate the reader
with many formulae nor to leave many of the basic mathematical results as
references in other work.  Therefore we present the necessary analytic tools
in an appendix.

\header{2. Patch Size, Wavenumbers, and Correlation Lengths}

	In this section we shall briefly discuss the relation between the
correlation length of a function on the hyperboloid and the wavenumbers which
characterize the mode decomposition.  Unlike the flat cosmology, in an open
cosmology there is a built-in spatial scale, namely the curvature scale
$\Rcurv$.  One consequence of this curvature scale is that we cannot
automatically use our Euclidean intuition, which should apply to
lengths below $\Rcurv$, to extrapolate to lengths comparable to and larger than
the curvature scale.

\subheader{Functions on a Euclidean space}

	Consider a function in 3-dimensional Euclidean space, $f(\bfx)$, which
consists of a positive ``patch'' of size $L$ and which falls off rapidly
outside of this patch.  An example of this would be a top-hat function:
$f(\bfx)=\Theta(L-|\bfx|)$, or a 3-dimensional Gaussian:
$f(\bfx)=\exp(-|\bfx|^2/(2L^2))$.  One could Fourier transform such a function
$$\widetilde{f}(\bfk)
       = {1\over(2\pi)^{3\over2}}\int d^3\bfx\,f(\bfx)\,e^{-i\bfk\ccdot\bfx},
        \eqno(2.1)$$
obtain the power spectrum by integrating the unit vector $\hatbfn$ over the
unit sphere
$$\calP_f(k)= {1\over4\pi}\int d^2\hatbfn\,|\widetilde{f}(k\hatbfn)|^2,
	\eqno(2.2)$$
and obtain the 2-point correlation function
$$\Xi_f(d)= 4\pi\int_0^\infty dk\,k^2{\sin kd\over kd}\,\calP_f(k).
	\eqno(2.3)$$
For either of the above functions which describe a patch of size $L$, we see
that most of the power,  i.e. the dominant contribution to the integral
$$4\pi\int_0^\infty dk\,k^2\,\calP_f(k)=\int d^3\bfx\,|f(\bfx)|^2
,\eqno(2.4)$$
is due to values of the wavenumber $k\sim1/L$.  As well, the correlation
function, $\Xi_f(d)$ , is roughly constant until $d\simgt L$ where it starts to
fall-off rapidly.  Thus, the Fourier-transform coefficients at wavenumber $k
\sim 1/L$ may be said to characterize the patch size and correlation length.

\subheader{Functions on a hyperboloid}

Now let us consider a similar function which consists of a ``patch'' of size
$L$ on a 3-dimensional hyperboloid.  A square-integrable function on a
hyperboloid, such as this patch, may be decomposed into scalar harmonics on the
hyperboloid, using a Mehler-Fock transform [see eqs~(A17-28)]. Certainly,
these harmonics are different from the usual Fourier transform harmonics.  For
instance, the eigenvalues, $-k^2$, of the Laplace-Beltrami operator acting on
the scalar harmonic eigenfunctions have a spectrum extending only to the range
$(-\infty,-1/\Rcurv^2]$ [see eq~(A17)].  If we identify $k$ as a wavenumber,
then there are no ``very long wavelength'' (eigen-)modes, i.e. with
$k<1/\Rcurv$.  As well, one can verify that eigenfunctions with eigenvalue
$-k^2$ really do vary with a typical length-scale given by $k^{-1}$.  Hence, no
eigenfunctions are smooth over a length scale much larger that the curvature
radius.  Thus if one has a function with a very large patch size, $L\gg\Rcurv$,
one cannot guess, as we do in Euclidean space, that it is composed mostly of
eigenmodes with $k\sim1/L$, since there are no such modes.  Rather, the way to
compose a function with a large patch size $L \gg \Rcurv$ out of eigenfunctions
which vary on length scales up to $\Rcurv$ is to carefully add eigenfunctions
with nearly the same patch size together but with opposite signs.  With the
right choice of coefficients one can, by delicate cancellation, construct a
function with an arbitrarily large patch size.

	To illustrate this we consider two examples of functions with very
large patch size and their Mehler-Fock transform.  It is convenient and
conventional to write the Mehler-Fock transform in terms of the variable
$\nu\equiv\sqrt{k^2\Rcurv^2-1}$ rather than the physical wavenumber $k$.  The
range of $\nu$ is $[0,\infty)$.  One should remember that $\nu=0$ corresponds
to a finite physical wavenumber $k=1/\Rcurv$ and not to an infinitely long
wavelength.

\subheader{Decomposition of a radial step function on the hyperboloid}

	Consider first a radial step function, i.e. a top-hat:
$$f(\chi,\theta,\phi)=\Theta(\chi-\chi_0)
,\eqno(2.5)$$
where $\chi$ is the radial coordinate measured in units of the curvature
radius, $\theta$ the polar angle, and $\phi$ the azimuthal angle [see eq~(A1)].
Since this function is spherically symmetric about the origin the transform
only contains the $(l,m)=(0,0)$ term. The Mehler-Fock transform of this
function is
$$\widetilde{f}_{lm}(\nu)={\sqrt{8}\over1+\nu^2}
               \bigl(\cosh\chi_0\sin\nu\chi_0-\nu\sinh\chi_0\cos\nu\chi_0\bigr)
               \delta_{l0}\delta_{m0}
.\eqno(2.6)$$
If the patch size, $\chi_0\Rcurv$, is much smaller than $\Rcurv$, i.e. if
$\chi_0\ll1$, then
$$\lim_{\chi_0\rightarrow0}\widetilde{f}_{lm}(\nu)
={\sqrt{8}\over\nu^2}\bigl(\sin\nu\chi_0-\nu\chi_0\cos\nu\chi_0\bigr)
                                                         \delta_{l0}\delta_{m0}
\eqno(2.7)$$
which is the usual transform of a top-hat in terms of spherical Bessel
functions $j_l$.  In this case most of the power is concentrated at
$\nu\chi_0\sim1$ or $kL\sim1$, just as our Euclidean intuition tells us. On the
other hand, if the patch size is much larger than the curvature radius, i.e.
$\chi_0\gg1$, then the transform of eq~(2.6) becomes oscillatory at $\nu\ll1$,
i.e $k\Rcurv=1$.  Now our Euclidean intuition tells us that for $\chi_0 \gg 1$
the dominant contribution to the transform should be due to that harmonic which
oscillates only on very large wavelengths.  This Euclidean intuition is
misleading when applied to the large-scale properties of functions on this
negatively-curved space.  Examine the behavior of eq~(2.6) for $\chi_0
\gg 1$, varying $\nu$. We see that the transform is dominated by contributions
from a continuous range of harmonics with $0<\nu \simlt 1$.
Thus, a single harmonic cannot identify or characterize this step function
on length scales larger than the horizon radius.  There is, however, nothing
mysterious about the behavior of these Mehler-Fock transforms.
The decomposition of this square integrable function which is constant on
length scales larger than the curvature radius requires an assembly of many
harmonics.

\subheader{Decomposition of a Gaussian function on the hyperboloid}

As a second illustration of the properties of functions and transforms on this
negatively curved space, consider a Gaussian function centered at the origin:
$$f(\chi,\theta,\phi)=\exp\left(-{1\over2}{\chi^2\over\chi_0^2}\right)
.\eqno(2.8)$$
The Mehler-Fock transform is
$$\widetilde{f}_{lm}(\nu)=2\sqrt{\pi}\chi_0\sin\nu\chi_0^2
 \exp\left(-{1\over2}(\nu^2 - 1)\chi_0^2\right) \delta_{l0}\delta_{m0}
.\eqno(2.9)$$
This function $\widetilde{f}_{lm}(\nu)$ is oscillatory in $\nu$ with angular
frequency $\chi_0^2$.  If $\chi_0 \ll 1$ then the amplitude of the transform is
exponentially damped as $\exp(-\nu^2\chi_0^2/2)$ before the first oscillation
in $\nu$.  In this case the transform closely approximates the flat-space
Fourier transform.  If the width of the Gaussian is much larger than the
curvature radius, $\chi_0\gg1$, then the function oscillates many times before
it is exponentially damped.

	The power spectrum of the Gaussian is simply related to square of
the Mehler-Fock transform of eq~(2.9) via eq~(A30).  Note that the power
spectrum of the Gaussian will also be oscillatory.  One may determine which
wavenumber contributes most of the power by considering the cumulative power as
a function of the physical wavenumber, $k$, i.e.
$$\calP_f^<(\nu)\equiv4\pi\int_0^\nu d\bar{\nu}\,\bar{\nu}^2\calP_f(\bar{\nu})
\eqno(2.10)$$
which for the Gaussian is given by
$$\calP_f^<(\nu)
=\pi\sqrt{\pi}\,\chi_0\,\left(e^{\chi_0^2}\erf\chi_0\nu
          -{1\over2}\left(\erf(\nu+i)\chi_0+\erf(\nu-i)\chi_0\right)\right)
.\eqno(2.11)$$
We plot the shape of this function versus $k$ for various values of $\chi_0$ in
figure 1.  We see that for a Gaussian with width much smaller than $\Rcurv$,
i.e. $\chi_0\ll1$, that most of the power is concentrated at
$k\sim1/(\chi_0\Rcurv)$, i.e. wavenumbers corresponding to the width of the
Gaussian.  However for Gaussians with width much greater than $\Rcurv$ the
power becomes concentrated at $k$'s very close to the minimum value:
$k=1/\Rcurv$, which is a wavenumber much greater than $1/(\chi_0\Rcurv)$.  As
we shall see this smaller length scale is also reflected in the correlation
function.


The notion that the Gaussian may be nearly constant on distances larger than
the curvature radius does not necessarily mean that the correlations in the
Gaussian extend, on average, over distances larger than the curvature scale.
To see this let us first calculate the correlation function of this Gaussian as
defined in (A29).  We obtain
$$\corrfuncSI{f}(d)
 ={\pi^2\chi_0^2e^{\chi^2_0}\over2\sinh d}\,
  \left[2\erf{d\over2\chi_0}-\erf{d-2\chi_0^2\over2\chi_0}
                            -\erf{d+2\chi_0^2\over2\chi_0}\right]
.\eqno(2.12)$$
which is illustrated in figure 2.  If $\chi_0\ll1$ then this correlation
function falls off for $d\simgt\chi_0$, while for $\chi_0\simgt1$ the
correlation function falls off for $d\simgt1$ no matter how large is $\chi_0$.

How can a function have a correlation length much smaller than the size of the
region over which the function extends?  To understand this one should recall
that $\corrfuncSI{f}$ gives the volume-weighted product of $f$'s at all pairs
of points separated by a given distance.  If the function has support on length
scales larger than the radius of curvature, $\chi_0\gg1$, then this weighted
product is dominated by contributions from the outer edges of the function
where $f$ falls off rapidly, i.e. it is dominated by $\chi\sim\chi_0$.  The
rapid variation on the outer edges overwhelms the contribution of $\chi<\chi_0$
due to the exponentially increasing volume at large radii in a hyperbolic
space.  Thus the correlation length of the 2-point function reflects the rapid
variation at the edges of the function $f$ and not the slower variation in the
center.

\header{3. Inequalities for Correlation Functions}

In this section we shall quantify the claim that large scale correlations are
suppressed in an open universe.  Consider two different moments of the power
spectrum of some scalar function $\Phi$
$$M_i=4\pi\int_0^\infty d\nu\,\nu^2\,K_i(\nu)\,\calP_\Phi(\nu)
\eqno(3.1)$$
for $i=1,2$. Here $\calP_\Phi(\nu)$ is defined by eq~(A30), and $K_i(\nu)$
is some arbitrary function of $\nu$. We know that $\calP_\Phi(\nu)\ge0$, so
that we have the inequalities
$${|M_2|\over M_1}\le\max_\nu {|K_2(\nu)|\over K_1(\nu)}
\qquad{\rm if}\qquad  K_1(\nu)>0 \qquad\qquad {\rm(Case\ A)}
\eqno(3.2)$$
or
$$\min_\nu{K_2(\nu)\over K_1(\nu)}\le{M_2\over M_1}
                                  \le \max_\nu{K_2(\nu)\over K_1(\nu)}
\qquad{\rm if}\qquad K_2(\nu), K_1(\nu)>0   \qquad {\rm(Case\ B)}
.\eqno(3.3)$$
Now, any sort of variance or 2-point function is a moment of the power spectrum
of the form of eq~(3.1).  From eq~(A29) we see that the weighting of the power
spectrum for the $\corrfuncSI{\Phi}(d)$ is $\sin(\nu d)/(\nu\sinh d)$.  For
$d=0$ this is just unity, which is positive definite.  Thus the ratio of any
other moment of the power spectrum to $\corrfuncSI{\Phi}(0)$ must obey the Case
A inequality. For example the ratio of the 2-point correlation at separation
$d$ to that at zero lag must obey
$${|\corrfuncSI{\Phi}(d)|\over\corrfuncSI{\Phi}(0)}
                        \le\max_\nu{|\sin\nu d|\over\nu\sinh d}={d\over\sinh d}
\eqno(3.4)$$
where the maximal ratio of the weighting function occurs at $\nu=0$.  We see
that the correlation function must fall off faster than $\sinh d/d$ which
itself falls off like $\exp(-d)$ for $d\simgt1$.

This quantifies the claim that for {\it any} square-integrable function, the
correlation length cannot greatly exceed the curvature radius.  For
non-square-integrable functions the correlation function $\corrfuncSI{\Phi}(d)$
is not defined so one must come up with a different definition of the
correlation function in order to obtain the correlation length.

	The functions which describe inhomogeneities in an open cosmology are
typically not square-integrable. Instead they are drawn from statistically
homogeneous and isotropic distributions. The expectation of bilinear moments of
these functions is given by the power spectrum of eq~(A32).  Thus, the
inequality of eq~(3.4) applies equally well to the case of homogeneous random
noise in an open universe, i.e. with $\corrfuncSI{\Phi}$ replaced by
$\corrfunc{\Phi}$.  In the case of cosmological inhomogeneities the correlation
function of scalar functions must fall off exponentially beyond the curvature
scale.  This can be compared to flat-space where the relevant inequality is
$|\corrfunc{\Phi}(d)|\le\corrfunc{\Phi}(0)$ which is trivially satisfied for
any distribution.  This inequality does not imply that there cannot be large
regions, much bigger than the curvature radius, in which a function is nearly
constant.  Rather it just says that on average - when averaging over many
realizations from a homogeneous, isotropic distribution - the field may vary
significantly over distances greater than the curvature radius.

The bound on the ratio of the correlation functions, eq~(3.4), indicates that
there is no way of smoothing functions in a statistically homogeneous way in
order to increase the correlation length beyond the curvature scale. For
example, one way of smoothing a function in a homogeneous and isotropic way is
to convolve the function with a spherically symmetric ``smoothing kernel''.  As
indicated by eq~(A35), convolution just multiplies the power spectrum by a
factor and this factor drops out in the ratios which give the inequalities of
eqs~(3.2-3).  Therefore eq~(3.4) applies to the smoothed function as well as
the unsmoothed function.

	The reason that smoothing is unable to increase the correlation length
is fairly easy to understand.  The smoothing kernel must be square integrable
in order for the smoothed function to retain a finite variance:
$\corrfunc{W*\Phi}(0)<\infty$.  As noted above square-integrability means that
the smoothing kernel, $W$, must fall off exponentially at some finite distance
from its ``center''.  Due to the exponential increase in volume at large
distances, it is this rapidly varying part of the kernel which will contribute
most to the variation of the smoothed function.  Thus the smoothing kernel
cannot effectively smooth things on scales greater than the curvature radius.

\header{4. Limits on $\Omega$}

	We have shown that the correlation function of scalar quantities which
describe the inhomogeneities in an open universe must fall off rapidly with
large spatial separation.  Can we use this property of the correlation function
to set limits on the curvature radius of our own universe and thus place limits
on $\Omega_0$?  Recall that we have only placed upper limits on the correlation
function and that this upper limit becomes smaller as one decreases the radius
of curvature, and therefore $\Omega_0$.  One can therefore only hope to place a
lower limit on $\Omega_0$ from the inequality we have derived.  If the
correlation function starts to fall off exponentially on some scale, it could
be that one has reached the curvature radius, or it could be for some other
reason.  However if we do not see an exponential fall-off then we have not
reached the curvature scale.

	We are proposing a {\it model independent} upper limit which only
depends on the assumption of isotropy and homogeneity.  If one had further
knowledge of the origin of the inhomogeneities which, say, fixed the power
spectrum as a function of $\Omega_0$ then one could hope to measure $\Omega_0$
directly from the correlation function.  Of course the predicted correlation in
any such model must obey the inequalities derived above.  So far in this paper
we have not assumed any model.

	One should also recall that the correlation functions we are dealing
with are expectation values under a distribution.  In most cases, i.e. for
ergodic processes, they should also give the correlation function for averages
over the entire space.  They do not give an indication of the distribution of
any quantity in a finite volume, i.e. how probable it is to obtain a set of
measurements.  This is determined by the statistics of the random process
generating the inhomogeneity.  Until one makes assumptions about this
distribution one cannot set any rigorous statistical limit on $\Omega_0$.
Nevertheless if $\Omega_0$ is sufficiently small and our observations are of a
sufficiently large volume  we may hope to have very good statistics.  This is
because, as we have demonstrated, the correlation length is limited to the
curvature radius and under a large class of statistical distributions this
would mean that roughly each volume of curvature radius is nearly statistically
independent.  Thus in our observational volume we would have a large number of
independent samples of the distribution and thus a representative sample of our
universe.

	The obvious application of the limit on the correlation function is to
the distribution of galaxies.  There are, however, at least three reasons why
this is not likely to be interesting.  First, as we see from equation (A14) the
actual size of the curvature radius is quite large, corresponding to redshifts
greater than $1.7$ and distances larger than $3000 h^{-1}\Mpc$.  This is a
distance much larger than it is practical to survey in the foreseeable future,
as galaxies at this distance are very faint.  Furthermore one would have to
understand the evolution of the luminosity function of galaxies before one
would be able to disentangle the evolution from the inhomogeneity.  Second, it
is well known that the galaxy correlation function decreases rapidly with
separation as far as is measured, up to about $100 h^{-1}\Mpc$ which is clearly
less than the curvature radius. While it is true that a simple extrapolation of
this fall-off would not match the exponential fall-off required in an open
universe, one would still have to measure the correlations where they were
extremely small in order to put an interesting limit on $\Omega_0$.  To do this
one would have to keep systematic errors to an extremely low level.  One might
also think that such a measurement would be nearly impossible because of ``shot
noise'', the statistical fluctuations caused by the finite number of galaxies.
This need not be the case since the shot noise, which after all is also
statistically homogeneous, would also be cutoff exponentially at the curvature
radius.  This decrease in shot noise is a result of the extremely large number
of pairs of galaxies at large separations due to the exponentially growing
volume element in a hyperbolic space.  This points to the third problem with
measuring $\Omega_0$ with galaxy correlations.  Namely that if one were able to
measure the galaxy distribution deep enough to notice the fall-off in
correlations, one would almost certainly notice the exponentially growing
volume element just by counting the number of galaxies.  This classical number
counting technique is almost certainly a better probe of $\Omega_0$ than the
correlation function.

	To get to large enough distances one might examine the correlations in
the locations of QSO's or their absorption systems.  These objects are much
easier to observe at large distances than galaxies, but they also suffer from
having an unknown evolution of their numbers and luminosity with redshift.  The
number density of observed QSO's is much less than that of galaxies so it is
more difficult to measure correlations on any scale, requiring a greater sample
volume for a significant result.  The number of Ly-$\alpha$ absorption systems
is quite large, especially at large redshifts, but one is limited to seeing
these systems only along the line-of-sight to QSO's.  In neither case is the
correlation function at separations of $\simgt3000 h^{-1}\Mpc$ likely to be
obtained in the near future.  A simple extrapolation of correlations on smaller
scales would suggest that the amplitude of inhomogeneity on these scales would
be very small and difficult to measure.

	An easier way to get to very large separations is to look at the MBR
with which we can see back to the last-scattering epoch at a redshift
$Z_{ls}\sim10-1000$.  Recently significant correlations in MBR temperature
anisotropies have been measured using the COBE satellite [6].  Such large angle
measurements are already sampling scales much larger than the curvature radius
even for moderately small $\Omega_0$.  The suppression of spatial correlations
at large separations should have the greatest effect on these large
angular-scale MBR measurements such as COBE.  We shall now proceed to study how
angular correlations of the MBR temperature anisotropies are effected by
curvature. Placing the observer at the origin ($\chi=0$) and assuming small
scalar deviations from homogeneity, the MBR anisotropy on large angular scales
is given by
$${\Delta T\over T}(\theta,\phi)
    ={1\over4}\delta_\gamma(\etaobs-\etals,\theta,\phi,\etals)
               +\Phi(\etaobs-\etals,\theta,\phi)
               +2\int d\eta\,\dot{\Phi}(\etaobs-\eta,\theta,\phi,\eta)
               -v_\chi(\etaobs-\etals,\theta,\phi)
\eqno(4.1)$$
where $\delta_\gamma$ gives the fractional change in the photon density and
$\Phi$ gives the fluctuation in the Newtonian potential, $\dot{\Phi}$ is its
derivative with respect to $\eta$, and $v_\chi$ is the radially directed
velocity of the photons.  The quantity $\etaobs-\etals$ gives the conformal
distance to the surface of last scattering, and hence the $\chi$ coordinate of
the points on the surface of last scattering.

	Let us now consider the relative importance of the terms contributing
to the MBR anisotropy in equation (4.1).  In standard flat cosmology the
Newtonian potential is constant in time so the 3rd, ``integrated Sachs-Wolfe'',
term in equation (4.1) is very small.  We also ignore the 4th, ``Doppler'',
term in equation (4.1) which is not likely to be important on very large
angular scales. It is the 1st, ``density'', and 2nd, ``potential'' terms which
are most important on large scales.  Whichever term dominates depends on the
nature of the perturbations.  For non-adiabatic fluctuations, deviations in the
photon-to-baryon ratio will {\it usually} lead to photon density fluctuations
which dominate the potential fluctuations.  For adiabatic fluctuations the
density fluctuations nearly cancel the potential fluctuations leading to a net
anisotropy of only ${1\over3}\Phi$. However in an open universe the
gravitational potential will decay and the integrated Sachs-Wolfe effect will
become important.  Below we show that if $\Omega_0$ is small, the large-scale
correlations in the density and potential anisotropies are strongly suppressed,
while the integrated Sachs-Wolfe anisotropies are less strongly suppressed.
For this reason we can expect that for small $\Omega_0$ that the large-scale
anisotropies will be dominated by the integrated Sachs-Wolfe effect.

  	We will not consider tensor and vector fields in this work, considering
only the cosmological perturbations described by scalar quantities.  Whereas it
is certain that scalar quantities such as density and gravitational potential
fluctuations play a significant role in the observed MBR anisotropy, the
relative importance of tensor and vector fields is empirically rather
uncertain, and can vary greatly between different cosmological models of
structure formation.  We expect that the techniques used in this paper may be
applied to vector and tensor modes and that one would obtain similar limits on
$\Omega_0$ from such calculations.

	One might also expect the suppression of large-angle correlations of
the  polarization of the MBR.  Polarization provides certain advantages over
temperature anisotropies in obtaining limits on $\Omega_0$.  Firstly because
the MBR polarization, generated by electron-photon scattering, is confined to
the surface of last scattering, while MBR anisotropy may be generated at
smaller redshifts via the integrated Sachs-Wolfe effect.  We will see below
that the integrated Sachs-Wolfe anisotropies yield much less stringent limits
on  $\Omega_0$ than those generated on the surface of last scattering.
Secondly there is no need to subtract off the monopole or dipole components of
polarization, they are zero.  Such subtractions in the case of anisotropies
lead to ``artificial'' correlations at large angular separations.  Of course
the main disadvantage of MBR polarization is that it has not yet been
detected.  Furthermore it is likely to be at a much smaller amplitude than
temperature anisotropies, and thus more difficult to measure accurately.  We
will not consider MBR polarization further in this paper.

\subheader{Anisotropies from the Surface of Last Scattering}

	Let us first consider the MBR anisotropies generated at the
surface-of-last-scattering, i.e. by ignoring the integrated Sachs-Wolfe
contribution. Thus we consider the density and potential contributions to the
anisotropy of equation (4.1).  The 2-point angular correlation function of this
component of the anisotropy can be easily written in terms of,
$\corrfunc{{1\over4}\delta+\Phi}$, the spatial 2-point correlation function of
the scalar field ${1\over4}\delta+\Phi$ or its power spectrum,
$P_{\s{1\over4}\delta+\Phi}$, i.e.
$$C_{\rm ls}(\psi)=\corrfunc{{1\over4}\delta+\Phi}
                          \left(d(\etaobs-\etals,\etaobs-\etals,\psi)\right)
                  =4\pi\int d\nu \nu\,P_{\s{1\over4}\delta+\Phi}(\nu)
                     {  \sin \nu d(\etaobs-\etals,\etaobs-\etals,\psi)
                      \over\sinh d(\etaobs-\etals,\etaobs-\etals,\psi)}
\eqno(4.2)$$
where $\psi$ is the angular separation and $d(\chi_1,\chi_2,\psi)$ defined in
equation (A3) gives the comoving distance between the two points on the surface
of last scattering an angle $\psi$ apart.  As computed in the previous section,
the inequality of equation (3.4) when applied to the intrinsic anisotropy gives
$${|C_{\rm ls}(\psi)|\over C_{\rm ls}(0)}
 \le{ d(\etaobs-\etals,\etaobs-\etals,\psi)\over
\sinh d(\etaobs-\etals,\etaobs-\etals,\psi)}
.\eqno(4.3)$$
Using equation (A13) to calculate $\etaobs-\etals$ as a function of $\Omega_0$
we have plotted the upper limit of equation (4.3) in figure 3 for various
values of $\Omega_0$. We see that the required fall-off in the correlation
function is very striking even for moderately small $\Omega_0$.


	  One cannot compare the curves of figure 3 directly to $C(\psi)$ of
the MBR anisotropy observed by COBE primarily because MBR anisotropy
experiments do not measure the intrinsic anisotropy on the surface of last
scattering but rather the sum of this with many other effects.  Furthermore MBR
anisotropy experiments such as COBE do not measure the 2-point function of
the anisotropy on the sky.  Instead the anisotropy is convolved with the beam
pattern of the experiment.  The beam patterns are not sensitive at all to the
monopole component of the anisotropy pattern and the dipole component of the
anisotropy pattern is explicitly subtracted.  We have seen above that the
limit of equation (3.4) is equally valid before and after 3-dimensional
convolution in a hyperbolic space. However convolutions on the sky, which are
intrinsically anisotropic, will invalidate this formula.  Nevertheless we can
use the inequalities of equation (3.2-3) to come up with alternative
inequalities which apply to the convolved anisotropy with monopole and dipole
subtraction.

\subheader{Effects of Beam Smearing}


	Here we demonstrate the effects of Gaussian convolution in an attempt
to address the problem of fluctuations on the surface
of last scattering more realistically. We convolve the anisotropy pattern
on the sky with a Gaussian window function of the form
$$f(\psi;\sigma)={\exp{\cos\psi\over\sigma^2}
                  \over4\pi\sigma^2 \sinh{1\over\sigma^2}}
\eqno(4.4)$$
where the width of the Gaussian may also written in terms of its FWHM,
$\theta_{\rm FWHM}$, given by
$$\sigma=\sqrt{1-\cos{1\over2}\theta_{\rm FWHM}\over\ln2}
.\eqno(4.5)$$
Applying this smoothing function to the anisotropy pattern, we find the
smoothed correlation function is
$$C_{\rm ls}(\psi;\sigma)=
{1\over16}\int d\nu  |\delta_{rms}(\nu)|^2
\left(\sigma^2\sinh{1\over\sigma^2}\right)^{-2}
\sum_{J=0}^{\infty} {2 J + 1 \over 4 \pi}
|\Pi^J_\nu(\etaobs-\etals) i_J({1\over\sigma^2})|^2 {\rm P}_J(\cos\psi).
\eqno(4.6)$$
Here, $i_J(x)=i^J j_J(i x)$, is a modified spherical Bessel function and
${\rm P}_J(x)$ a Legendre polynomial. Due to the simple form of the Gaussian
smoothing function, we may easily subtract the contributions of any multipole
moment of the anisotropy pattern by eliminating that term from the sum over
$J$. We shall adopt the notation $C_{\rm ls}(\psi;\sigma)_l$ to indicate that
the sum over $J$ begins with the $J=l$ term.

	Consider the zero-lag correlation function:
$$C_{\rm ls}(0;\sigma)_l= {1 \over 16} \int d\nu  |\delta_{rms}(\nu)|^2
\left(\sigma^2 \sinh{1\over\sigma^2}\right)^{-2}
\sum_{J=l}^\infty{2J+1\over4\pi}
                        |\Pi^J_\nu(\etaobs-\etals)\,i_J({1\over\sigma^2})|^2.
\eqno(4.7)$$
Comparing equation (3.1) with (4.7), we see that the term in the above equation
which corresponds to $K(\nu,x)$ is positive for any smoothing width $\sigma$.
Therefore, as in equation (3.3), we may place an upper {\it and} lower bound on
the ratio of two variances with different smoothing widths. In particular
$$ \min_\nu r(\sigma_1,\sigma_2,\nu)_l \le
 {C_{\rm ls}(0;\sigma_1)_l \over C_{\rm ls}(0;\sigma_2)_l}
\le \max_\nu r(\sigma_1,\sigma_2,\nu)_l
\eqno(4.8)$$
where the ratio, $r(\sigma_1,\sigma_2,\nu)_l$, is given by
$$r(\sigma_1,\sigma_2,\nu)_l=
\left[{\sigma_2^2\sinh{1\over\sigma_2^2}\over
       \sigma_1^2\sinh{1\over\sigma_1^2}     }\right]^2
{\sum_{J=l}^\infty(2J+1)|\Pi^J_\nu(\etaobs-\etals)\,i_J({1\over\sigma_1^2})|^2
 \over
 \sum_{J=l}^\infty(2J+1)|\Pi^J_\nu(\etaobs-\etals)\,i_J({1\over\sigma_2^2})|^2}
.\eqno(4.9)$$
The minimum and maximum values of $r(\sigma_1,\sigma_2,\nu)_l$ may be easily
computed numerically.

In figure 4 we present the upper and lower bounds on the ratio of MBR variance
with $7^\circ$ and $90^\circ$ smoothing, having subtracted the monopole and
dipole moments. We have computed the same ratio
$C_{\rm ls}(0;\sigma(7^\circ))_2/C_{\rm ls}(0;\sigma(90^\circ))_2$ from the
2-year results of the COBE DMR experiment [6]. Comparing the observed results
with predicted bounds, we see that $\Omega_0$, the present-day cosmological
density parameter,  is only weakly restricted.

\subheader{Adiabatic Fluctuations and the MBR}

	The above models of anisotropy were not completely realistic since the
integrated Sachs-Wolfe effect was ignored.  The relative importance of the
integrated Sachs-Wolfe effect depends somewhat on the nature of the
perturbations.  Here we shall consider adiabatic fluctuations including the the
density, potential, and integrated Sachs-Wolfe contributions to the
anisotropies.  Following ref~[4] and assuming an adiabatic growing mode in a
dust-dominated open universe we find that the MBR anisotropy is given by
$${\Delta T\over T}(\theta,\phi)
={1\over3}\Phi(\etaobs,\theta,\phi)
 +2\int_0^\etaobs\rmd\eta\,\Phi_0(\etaobs-\eta,\theta,\phi)\,F'(\eta)
\eqno(4.10)$$
where the time evolution of the potential is given by the function $F$ (see
equation (A16)).  The angular correlation function may thus be written as
$$\eqalign{
C_{\rm ad}(\psi)={1\over9}\corrfunc{\Phi}(d(\etaobs,\etaobs,\psi))
         &+{2\over9}\int_0^\etaobs\rmd\eta\,F'(\eta)
                              \corrfunc{\Phi}(d(\etaobs,\etaobs-\eta,\psi)) \cr
         &+4\int_0^\etaobs\rmd\eta_1\,F'(\eta_1)
           \int_0^\etaobs\rmd\eta_2\,F'(\eta_2)
                     \corrfunc{\Phi}(d(\etaobs-\eta_1,\etaobs-\eta_2,\psi))
           }\eqno(4.11)$$
or
$$C_{\rm ad}(\psi)
        =4\pi\int_0^\infty\rmd\nu\,\nu\,\calP_{\Phi}(\nu)\,K_{\rm ad}(\nu,\psi)
.\eqno(4.12)$$
Here the kernel is
$$\eqalign{K_{\rm ad}(\nu,\psi)
=&{1\over9}{\sin \nu d(\etaobs,\etaobs,\psi)\over\nu\,
            \sinh    d(\etaobs,\etaobs,\psi)}                               \cr
+&{2\over3}\int_0^\etaobs\rmd\eta\,F'(\eta)
           {\sin \nu d(\etaobs,\etaobs-\eta,\psi)\over\nu\,
            \sinh    d(\etaobs,\etaobs-\eta,\psi)}                          \cr
+&4\int_0^\etaobs\rmd\eta_1\,F'(\eta_1)
   \int_0^\etaobs\rmd\eta_2\,F'(\eta_2)\,
           {\sin \nu d(\etaobs-\eta_1,\etaobs-\eta_2,\psi)\over\nu\,
            \sinh    d(\etaobs-\eta_1,\etaobs-\eta_2,\psi)}                 \cr
           }\eqno(4.13)$$
which at zero lag is
$$K_{\rm ad}(\nu,0)
={1\over9}
 +{2\over3}\int_0^\etaobs\rmd\eta\,F'(\eta){\sin\nu\eta\over\nu\,\sinh\eta}
 +4\int_0^\etaobs\rmd\eta_1\,F'(\eta_1)
   \int_0^\etaobs\rmd\eta_2\,F'(\eta_2)\,
                     {\sin\nu(\eta_2-\eta_1)\over\nu\,\sinh(\eta_2-\eta_1)}
.\eqno(4.14)$$
To find the maximum allowed values of $C_{\rm ad}(\psi)/C_{\rm ad}(0)$ we
should maximize the ratio of the integrands $K_{\rm ad}(\nu,\psi)/K_{\rm
ad}(\nu,0)$ with respect to $\nu$.  This has been done numerically and the
results are displayed in figure 5 for $\Omega_0$ in the range $[0.01,0.1]$.
Much smaller values of $\Omega_0$ are required to produce the same suppressions
exhibited in figure~3, and therefore the limits on $\Omega_0$ for adiabatic
perturbations will be much less stringent than found in the case where the
anisotropies are generated solely on the surface of last scattering.


	The reason that the bound on $\Omega_0$ using
$C_{\rm ad}(\psi)/C_{\rm ad}(0)$ is weaker than the bound obtained using
$C_{\rm ls}(\psi)/C_{\rm ls}(0)$ is not hard to understand.  All of the
contributions to $C_{\rm ls}(\psi)$ are at the relatively large comoving
distance $\etaobs-\etals$, which for even moderately low values of $\Omega_0$
is much larger than a curvature radius away.  Unless the angular separation,
$\psi$, is very small two points on the sky will be several curvature radii
away from each other.  In this case $d\simgt1$ so that correlations will be
suppressed expolnentially. However for the integrated Sachs-Wolfe effect the
anisotropies are generated at much closer distances to the observer.  In fact
one needs $\Omega_0\simlt0.1$ before most of the integrated Sachs-Wolfe
anisotropy is generated at more than a curvature radii away.  Thus for a given
angular separation, $\psi$, the anisotropies along different lines of sight are
generated at relatively small spatial separations and the $d/\sinh d$
suppression of spatial correlations is smaller.  Thus for adiabatic
fluctuations the suppression of correlations at a given angular scale is
reduced because one is sampling a much smaller spatial separation than for
anisotropies generated on the surface-of-last-scattering.

	One should also notice that in all cases the upper bound on the
angular correlation function stops decreasing rapidly at some moderate angle.
This can be understood in terms of the the peculiar geometrical property of
hyperbolic spaces.  Namely that the spatial separation of two points an angle
$\psi$ apart as seen from an observer will have only a weak dependence on the
angle if the distance from the observer is much larger than $\Rcurv$ (see
equation (A4)).  Thus the $d/\sinh d$ suppression of spatial correlations
saturates at
some moderate angular scale and does not cause much further suppression at
larger angular separations.

\header{5. Anisotropies Without Power Spectra or Square-Integrable Functions}

So far we have considered global perturbations which are described by a power
spectrum or local perturbations described by square-integrable functions.  We
have shown that for such perturbations, the correlations necessarily fall off
at the curvature radius.  For homogeneous Gaussian random noise in terms of the
usual basis functions, this behavior leads to functions with support on length
scales which are never much bigger than the curvature radius.  It is certainly
feasible to construct functions with a patch size much larger than the
curvature radius, although these functions would not be very likely to arise
from Gaussian random noise.  As long as such functions are square-integrable
the correlation function will still fall off rapidly at the curvature radius.
This is due to the fact that the ``edges'' of the regions of positive or
negative support, which must vary rapidly in order to insure that the function
is square-integrable, dominate the average.  Thus, we are lead to the question
of whether there exist non-square-integrable functions which have correlation
lengths much larger than the curvature radius and whether these are relevant to
cosmology [7].

In this section, we shall examine two non-square-integrable functions with a
correlation length which is much larger than the curvature radius.  We shall
consider two examples of such a function, and show that if these functions
represent MBR temperature as emitted from the surface-of-last-scattering, that
the resultant anisotropy is still suppressed on large angular scales if the
surface-of-last-scattering is much further from the observer than the curvature
radius.  As we shall see the amount of suppression need not be as large as we
have found so far.

\subheader{A Step Function Along the Central Plane}

Consider a function which takes one value in one half of the space and
the opposite value in the other half.  In particular consider the function
$$T(\chi,\theta,\phi)=\sgn({\pi\over2}-\theta)
\eqno(5.1)$$
This clearly splits the space $H^3$ in half since there is a reflection
symmetry about $\theta={\pi\over2}$.  One might argue that this function is
just the limiting form of a very large top-hat function centered at infinity,
which is square-integrable, and therefore should also have rapidly falling
correlations on average. For any pair of points with separation $r$, however,
there are infinitely more pairs for which both points lie to one side of the
plane than pairs which straddle the plane. Therefore, one would also sensibly
argue that the 2-point correlation function is really just $+1$, independent of
separation.  (A simpler example of such a function is just a constant function,
whose 2-point correlation function is independent of separation.  This clearly
violates the limits which apply to functions described by a power spectrum. We
have chosen not to consider such a function, since it involves no
inhomogeneity.)

We shall consider the scenario in which equation (5.1), through some process,
determines the temperature of photons as they are emitted at the time of
last-scattering. That is, the temperature of the photons observed on the
celestial sphere will take one of two values, $+1$ for those photons emitted
at points with $\theta<{\pi\over2}$ and $-1$ for the rest.  The boundary
between these two temperatures is given by the intersection of the sphere of
radius $\etaobs$ centered on the observer and the plane $\theta={\pi\over2}$.
An observer located within a distance $\etaobs$ of the plane will see a
disk-like temperature anisotropy in the shape of a circle while an observer
located further than a distance $\etaobs$ of the plane will see no temperature
inhomogeneity on the sky. Now we will determine the distribution of the sizes
of disks as seen by observers.

We would like to determine the measure for the distribution of disk sizes as
seen by observers along the plane. Let us begin by defining  the distance from
an observer to the central plane, $D_\rmc$, to be the shortest distance
connecting the observer to any point on the central plane.  For an observer
located at coordinates $(\chi,\theta,\phi)$,
$$D_\rmc=\cosh^{-1}\bigl(\sqrt{1+\sinh^2\chi\cos^2\theta}\bigr)
\eqno(5.2)$$
gives the distance to the central plane.

The disk size is just a function of the distance of the observer from the
central plane, $D_\rmc$.  Therefore it suffices to determine the distribution
of $D_\rmc$.  We would like to determine the volume-weighted average of
$D_\rmc$.  Unfortunately this is not completely well defined since the volume
of points with $D_\rmc$ in some interval is infinite. To better define how to
weight the different distances one can make use of the fact that there is an
isometry corresponding to translations along the plane.  Combining these
translations with reflections about the central plane we may find that all
points a given distance $D_\rmc$ from the central plane are isometric.  By
requiring that the weighting scheme have the same symmetries unambiguously
determines the distribution of $D_\rmc$.  To determine this distribution,
consider the translations in the ``$x$'' and ``$y$'' directions, generated by
the two Killing vectors which map the central plane into itself (see equation
A10)
$$\eqalign{
\xi^\rmx_i=&(\cos\phi\,\sin\theta
             , \sinh\chi\,\cosh\chi\,\cos\theta\,\cos\phi
             ,-\sinh\chi\,\cosh\chi\,\sin\theta\,\sin\phi)                  \cr
\xi^\rmy_j=&(\sin\phi\,\sin\theta
             ,\sinh\chi\,\cosh\chi\,\cos\theta\,\sin\phi
             ,\sinh\chi\,\cosh\chi\,\sin\theta\,\cos\phi)
          }.\eqno(5.3)$$
The proper invariant weight, $w(D_\rmc)$ is given by the requirement that
$$\xi^\rmx_{[i}\xi^\rmy_j D_{\rmc,k]}=w(D_\rmc) \epsilon_{ijk}
\eqno(5.4)$$
where $[ijk]$ indicates antisymmetrization and $\epsilon_{ijk}$ is the
completely anti-symmetric symbol, such that $\epsilon_{123} = \sqrt{\gamma}$.
This is equivalent to requiring the volume swept out by an infinitesimal
translation in the two directions and by an infinitesimal change in $D_\rmc$
satisfy
$$w(D_\rmc)\propto{\delta V\over\delta\lambda_\rmx\delta\lambda_y\delta D_\rmc}
.\eqno(5.5)$$
Along the $\theta=0$ axis, we find that $w(D_\rmc) = {1 \over 6}\cosh^2D_\rmc$.
Interpreting this weight as a probability distribution we find
$$p(D_\rmc)\,dD_\rmc\propto \cosh^2 D_\rmc\,dD_\rmc
.\eqno(5.6)$$
This measure is not normalizable, but since we are only interested in the range
of values $D_\rmc < \etaobs$, we can still obtain sensible results.  The
cumulative probability that $D_\rmc$ is less than some value is
$$P_<(D_\rmc)\propto2D_\rmc+\sinh2D_\rmc
.\eqno(5.7)$$
For an observer a distance $D_\rmc$ away from the plane the angular radius
$\theta_\rmd$ subtended by the disk of different temperature is given by
$$\theta_\rmd = \cosh^{-1}\bigl( \tanh D_\rmc \coth \etaobs \bigr)
.\eqno(5.8)$$
So we find that the cumulative probability that $\theta_\rmd$ is larger than
some value is
$$P_>(\theta_\rmd)
\propto 2\tanh^{-1}\left(\cos\theta\,\tanh\etaobs\right)
        +\sinh2\tanh^{-1}\left(\cos\theta\,\tanh\etaobs\right)
,\eqno(5.9)$$
and the distribution function is
$$p(\theta_\rmd)={dP>(\theta_\rmd)\over d\theta_\rmd}\propto
{4\over2\etaobs+\sinh2\etaobs}
      {\sin\theta_\rmd\tanh\etaobs\over(1-\cos^2\theta_\rmd\,\tanh^2\etaobs)^2}
.\eqno(5.10)$$
Here we have chosen the normalization so that the integral from 0 to
${\pi\over2}$ is unity.  If $\etaobs\gg1$ then this distribution is strongly
peaked near $\theta\rightarrow0$, i.e. very small disk sizes.


In figure~6, we have plotted the probability of observing a patch of angular
size $\theta_\rmd$, for values of $\etaobs={1\over2}\Rcurv,\,\Rcurv,\,2\Rcurv$.
The peak in the distribution shifts to small angular patch sizes with
increasing $\etaobs$, according to the formula $\theta_{\rm peak}
=\min[\arcsin(1/(\sqrt{3} \sinh \etaobs)),{\pi/2}]$.  For a dust-dominated
expansion, relating $\etaobs$ to the surface of last scattering at a redshift
of $Z_{\rm ls} = 1000$, the three curves in figure~6 correspond to values of
the cosmological density parameter $\Omega_0 = 0.9, 0.8, 0.4$ respectively.
Thus, we find that for a cosmological model in which temperature anisotropy is
determined by the non-integrable function with correlation length longer than
the radius of curvature, given by eq~(5.1), then the spatial curvature {\it
still} serves to suppress the angular size of disks of temperature
inhomogeneities on the celestial sphere. As well, in such a toy model,
observation of a large disk of temperature inhomogeneity on the celestial
sphere may be translated into a lower bound on $\Omega_0$, the cosmological
density parameter.

\subheader{A Radial Gradient}

	Another class of functions one might consider are those which are
spherically symmetric about the origin, i.e.
$$T(\chi,\theta,\phi)=F(\chi)
.\eqno(5.11)$$
As above we shall suppose that $T(\chi,\theta,\phi)$, through some process,
determines the temperature of photons as they are emitted at the time of
last-scattering.  Clearly all observers will see an axisymmetric temperature
pattern with the axis of symmetry being the direction from the observer to the
center of symmetry.  If we place the observer a distance $\Robs$ from the
center of symmetry and draw a sphere of radius $\etaobs$ centered about the
observer, one finds that the distance, $R$, from the center of symmetry to a
point on this sphere is given by (see eq~(A3))
$$\cosh R(\alpha)=\cosh\Robs\,\cosh\etaobs-\sinh\Robs\,\sinh\etaobs\,\cos\alpha
,\eqno(5.12)$$
where $\alpha$ is the angle on the observers sky between the direction to this
point on the sphere and the direction to the center of symmetry. All distances
are measured in units of $\Rcurv$.  Here $\etaobs$  approximates the distance
to the surface-of-last-scattering so the temperature pattern is given by
$${\Delta T\over T}(\alpha)=F(R(\alpha))
,\eqno(5.13)$$
If $T$ is not square-integrable then it is difficult to define an average over
observers in order to conclude whether the anisotropies more often come from
large or small angular scales.   However any sensible average would give much
weight to observers very far from the center, and one can show that for
observers sufficiently far from the center the anisotropy pattern becomes
independent of this distance.  This is illustrated by the relationship
$$R(\alpha)-\Robs\approx\ln\left|\cosh\etaobs-\sinh\etaobs\,\cos\alpha\right|
\qquad \Robs\gg1,\,\etaobs
,\eqno(5.14)$$
One would think that to make large angle anisotropies one should take $F(R)$ to
be slowly varying.  If we take $F(R)$ to be sufficiently slowly varying that a
1st order Taylor series is a good approximation in the interval
$R\in[\Robs-\etaobs,\Robs+\etaobs]$, then
$${\Delta T\over T}(\alpha)-\overline{\Delta T\over T}=F'(\Robs)\,
\left(\ln\left|\cosh\etaobs-\sinh\etaobs\,\cos\alpha\right|
      +1-\etaobs{\cosh\etaobs\over\sinh\etaobs}\right)
.\eqno(5.15)$$
Here we have explicitly subtracted the mean temperature averaged over the sky
since this contributes isotropy not anisotropy. In figure~7 we plot this
temperature pattern for various values of $\etaobs$. Again we find that as we
increase $\etaobs$ past the curvature radius that the temperature anisotropy
becomes more and more concentrated on smaller angular scales.


	According to eq~(4.3) and figure~3 all square-integrable functions
will have a correlation function that will approach a $\delta$-function at zero
lag in the limit $\etaobs\gg1$.  We will now show that if $F$ is not
square-integrable then this limiting form may be evaded.  In the limit that
$\etaobs$ becomes very large, but still much smaller than the coherence length
of $F$, we obtain the limiting form of eq~(5.15)
$${\Delta T\over T}(\alpha)-\overline{\Delta T\over T}=F'(\Robs)\,
\left(\ln\left|{1-\cos\alpha\over2}\right|+1\right)
\qquad \Robs\gg\etaobs\gg1
.\eqno(5.16)$$
This exhibits a divergent hot or cold spot at $\alpha=0$, but note that the
spot is only logarithmically divergent.   If $0<\Omega_0\ll1$ and $F(\Robs)$
was some power law at large $\Robs$ then the observers at large distance from
the origin would see a pattern like that described by eq~(5.16).  Note that
such a power law would ensure that $F$ was not square-integrable and thus might
evade the limits set in \S4. Since essentially all the volume is at large
distance one could reasonably claim that the shape of the volume averaged
angular correlation function for the anisotropy is just given by the shape of
the angular correlation function of the pattern described by eq~(5.16). Since
the logarithm is an integrable singularity it is clear that
$C_{\rm ls}(\theta;0)_1$ is finite at $\theta=0$ and most other values of
$\theta$.  Thus if $F$ is asymptotically a power law then the correlation
function does not obey the limit of eq~(4.3).

	In this section we have examined two spatial temperature distributions
with very large scale spatial coherence.  In both cases the anisotropy patterns
that they induce are increasingly small angular scales as one increases the
radius of the surface-of-last-scattering past the curvature radius, i.e. as one
decreases $\Omega_0$ well below unity.  This illustrates that there is no
simple correspondence between large spatial coherence and large angular
coherence in an open universe.  It is not clear whether there are any
temperature configurations in which most observers would see only dipole
(and/or quadrupole) temperature anisotropy patterns if $\Omega_0$ is very
small.  This is in contrast to Euclidean space where it is quite easy to push
most of the anisotropies to the lowest multipoles by putting most of the
inhomogeneity on super-horizon scales.  In the case of radial gradients it
was shown that only a finite fraction of the anisotropy is pushed to large
multipole moments as $\Omega_0$ is decreased.  Thus it is possible for the
large angle correlations to remain large even if $\Omega_0$ is small.  A
question we have not fully answered is whether this is ever likely to occur in
a cosmological setting.  Clearly this will never occur in any distribution
described by a power spectrum.

	While we have considered only anisotropies generated on the
surface-of-last-scattering in this section we would expect qualitatively
similar results for anisotropies generated by adiabatic fluctuations.  One
would have to go to significantly larger values of $\etaobs$ in order to
obtain the same level of large-angle suppression for adiabatic inhomogeneities.

\header{6. Conclusion}

In this work we have examined certain statistical properties of functions in a
hyperbolic space, with applications to density fluctuations and MBR
anisotropies in an open universe.  We have shown that large-scale correlations
are exponentially suppressed for separations above the curvature scale in an
open universe. This suppression is generic to the open cosmology, rather than
being a feature of a particular model of inhomogeneities in an open cosmology.
We have further shown how the observed presence of large angular scale
anisotropy may be used to give a lower bound on the cosmological density
parameter $\Omega_0$ which is independent of the shape of the spectrum of
perturbations.  We have done this by considering two generic mechanisms for the
generation of MBR anisotropy, i.e.  adiabatic or last-scattering-surface
fluctuations, although we expect that similar sorts of limits may be obtained
for any combination primordial adiabatic and isocurvature fluctuations.

	The best limits on $\Omega_0$ come from the largest angular scale
anisotropies, such as measured by COBE [6].  We have not made a detailed
comparison of the suppression of large-scale anisotropies due to curvature with
the COBE data.  Effects such as instrumental noise and cosmic variance have not
been considered and we have not tried to find the most stringent limit on
$\Omega_0$ by searching for an optimal statistic. Furthermore any formal limit
would require assumptions about the ``statistics'' of the anisotropies.  The
comparison with COBE in figure~4 does indicate that even with extremely
optimistic assumptions only a relatively weak lower limit, $\Omega_0\simgt0.05$
can be obtained.  Considering adiabatic fluctuations while including
instrumental noise and a reasonable model for cosmic variance would yield a
significantly less stringent limit.  It seems likely that the lower limit on
$\Omega_0$ from statistics of anisotropies will never yield a more stringent
limit than is obtained from other methods.  Limits from statistics are,
however, a qualitatively different types of model-independent limit than others
that have been used.

	We have illustrated the suppression of large-scale correlations with
two classes of functions, namely square-integrable functions, and homogeneous
and isotropic ensembles of functions described by a power spectrum in the usual
way.  The former provides an instructive example while the latter might
describe the inhomogeneities in our own universe.  These two types of functions
were chosen for a very practical reason, namely that for these functions we
know how to compute averages to obtain 2-point correlation functions: a volume
average for the former and an ensemble average for the latter. It has suggested
by K. Gorski [7] that there may be distributions of cosmological
inhomogeneities which cannot be described by a power spectrum in the usual way,
and that such a distribution may not exhibit the suppression of large-scale
correlations found here.  This has yet to be shown.

	In \S5 we have considered spatial distributions which are not
square-integrable and unlike anything one is likely to find from a homogeneous
distribution of functions.  These functions can be said to have very large
spatial coherence, although this is difficult to quantify since the 2-point
correlation function as normally defined would not be finite. Nevertheless, in
spite of this large coherence we find that if these functions represented
temperatures that the typical anisotropies, i.e. for most observers, would have
significant power on very small angular scales if $\Omega_0\ll1$. It is the
geometry of the hyperbolic space that causes large spatial correlations lengths
to result in small angular correlation lengths for these configurations. For
the configurations examined in \S6 we find that the suppression of large-angle
anisotropies when $\Omega_0$ is small can be much less than the suppression
which is required for spatial distributions of temperatures describes by a
power spectrum.  It is therefore possible to find spatial distributions where
the volume-averaged anisotropy has large angular coherence when $\Omega_0$ is
small.  Whether such distributions are ever likely to arise in any model of
statistically homogeneous random noise is an unanswered question.  In any model
described by a power spectrum such configurations would essentially never
occur.

\noindent{\bf Acknowledgements:} Thanks to  Bruce Allen, Martin Bucher, Josh
Frieman, Kris Gorski, David Lyth,
Adi Nusser, Misao Sasaki, and Igor Tkachev for useful discussions. The work of
AS and RRC is supported by the NASA through grant number NAG-5-2788 (at
Fermilab). The work of RRC is funded by PPARC through grant number GR/H71550
(at Cambridge).
\vfill\eject

\header{Appendix: Tools for an open universe}

\subheader{Coordinates in an open universe}

The line element for a homogeneous, isotropic FRW spacetime with
spatial geometry of $H^3$, a 3-hyperboloid, is
$$d s^2=a^2(\eta)\left(-d\eta^2+d\chi^2
                   +\sinh^2\chi\,(d\theta^2+\sin^2\theta\,d\phi^2)\right)
\eqno(\rmA1)$$
where $a$ is the expansion scale factor, $\eta\in(0,\infty)$ is the conformal
time, and the spatial coordinates $\chi$, which gives the comoving radius from
the central point, $\theta$ which is the polar angle from an axis, and $\phi$
which is an azimuthal angle about this axis, have ranges $[0,\infty)$,
$[0,\pi]$, and $[0,2\pi]$, respectively.  Of course the choice of the central
point and the polar axis is arbitrary.  Units of length are carried by the
expansion scale factor. The spatial, $(\chi,\theta,\phi)$, part of this metric
we refer to as $\gamma_{ij}$ and its determinant as $\gamma$.  The curvature
of the spatial sections is $-1/a^2$ in physical units, which decrease with
time.  This means that in units of comoving curvature radius, $\Rcurv/a$ is
unity.  The
volume element of a spatial section is given by
$$d V=a^3\sinh^2\chi \sin\theta  d\chi d\theta d\phi,
\eqno(\rmA2)$$
so that the volume within a distance $\chi$ grows like $\sim e^{2\chi}$ for
$\chi$ larger than the curvature scale, $\chi \gg 1$. The comoving geodesic
distance, $d(\chi_1,\chi_2,\psi)$ between two points,
$(\chi_1,\theta_1,\phi_1)$ and $(\chi_2,\theta_2,\phi_2)$ is given by
$$\eqalign{\cosh d=&\cosh\chi_1 \cosh\chi_2-\sinh\chi_1 \sinh\chi_2 \cos\psi\cr
\cos\psi=&\cos\theta_1\cos\theta_2+\sin\theta_1\sin\theta_2\cos(\phi_1-\phi_2)
           }.\eqno(\rmA3)$$
Here $\psi$ is the angle between the direction to the two points as seen from
the origin.  Equation (A3) is simply the law of cosines for a hyperbolic
geometry. One curious property is that if the distance from the origin is much
greater than the curvature scale, then
$$\cosh d \approx \exp( \chi_1+\chi_2+\ln[\sin^2(\psi/2)] )
 \qquad \chi_1,\,\chi_2\gg1.
\eqno(\rmA4)$$
Hence, if the surface of last-scattering of MBR photons is much further from us
than the curvature radius then most pairs of points on this surface are
approximately at the same distance from each other with only a weak,
logarithmic, dependence on the angle between them.

\subheader{Isometries}

Here we may list the isometries of the 3-hyperboloid.  The group of isometries
of $H^3$ is $O(1,3)$, which has $6$ generators.  The isometries of the
3-hyperboloid may be specified by the 6 linearly independent Killing vector
fields on the manifold, i.e. vector fields satisfying Killing's equation
$$\xi^{i;j}_{(a)}+\xi^{j;i}_{(a)}=0 \qquad a=1,\ldots,6
.\eqno(\rmA5)$$
The Killing vectors generate diffeomorphisms of $H^3\rightarrow H^3$ of the
form
$$x^i\rightarrow X^i_{(a)}(\{x^j\},\lambda)
\eqno(\rmA6)$$
where $X^i_{(a)}(\bfx,\lambda)$ is a solution of the 1st order ordinary
differential equation
$${d\over d\lambda}X^i_{(a)}=\xi^{i}_{(a)}
\eqno(\rmA7)$$
with initial conditions
$$X^i_{(a)}(\{x^j\},0)=x^i
.\eqno(\rmA8)$$
Linear combinations of the diffeomorphisms may be constructed to form a
(sub-)group of isometries of the space.  Additional discrete symmetries may
lead to parts of the group which are disconnected from the identity.

The Killing vectors generating the 3 rotations, analogs to the rotations about
the ``$x$'', ``$y$'', and ``$z$'' axes in a Euclidean space, are respectively
$$\eqalign{
\xi^i_{(1)}=&(0,\sin\phi, {\cos\theta\over\sin\theta}\cos\phi)              \cr
\xi^i_{(2)}=&(0,\cos\phi,-{\cos\theta\over\sin\theta}\sin\phi)              \cr
\xi^i_{(3)}=&(0,0,1).                                                       \cr
           }\eqno(\rmA9)$$
These Killing vectors form the sub-group $O(3)$ with topology $S^2$. The 3
``translations''
$$\eqalign{
\xi^i_{(4)}=&(\cos\phi\,\sin\theta,
              {\cosh\chi\over\sinh\chi} \cos\phi\,   \cos\theta
            ,-{\cosh\chi\over\sinh\chi}{\sin\phi\over\sin\theta})           \cr
\xi^i_{(5)}=&(\sin\phi\,\sin\theta,
              {\cosh\chi\over\sinh\chi} \sin\phi\,   \cos\theta
             ,{\cosh\chi\over\sinh\chi}{\cos\phi\over\sin\theta})           \cr
\xi^i_{(6)}=&(\cos\theta         ,{\cosh\chi\over\sinh\chi}\sin\theta,0)    \cr
           }\eqno(\rmA10)$$
are analogs to the translations of the origin along the ``$x$'', ``$y$'', and
``$z$'' axes in a Euclidean space.

\subheader{Expansion Law}

The matter in our universe appears to be dominated by cold, pressureless dust.
In the absence of a cosmological constant, the cosmological density parameter
is given by
$$\Omega={8\pi G\rho\over3H^2}={1\over1+a} \qquad a={1-\Omega\over\Omega}
\eqno(\rmA11)$$
where $H$ is the Hubble constant and the normalization of the scale factor,
$a$, is chosen so that $a=1$ when $\Omega={1\over2}$. The evolution of the
scale factor follows as
$$a(\eta)=\sinh^2{\eta\over2}
.\eqno(\rmA12)$$
The horizon and the density parameter are related by
$$\eta=\ln{2-\Omega+\sqrt{(2-\Omega)^2-\Omega^2}\over\Omega}
.\eqno(\rmA13)$$
In physical units the curvature radius is given by
$$R_{\rm curv}={c\over H\sqrt{1-\Omega}}={3000\,h^{-1}\Mpc\over\sqrt{1-\Omega}}
  \qquad H=100h\,\km/\rms/\Mpc
.\eqno(\rmA14)$$
Even if $\Omega$ is very small we see that the curvature scale remains very
large.  Galaxies one curvature radius away have a redshift  greater than
$z\ge e-1\approx1.7$ and the lower limit is only approached for very small
$\Omega$.  Thus, the surface of last scattering, at a redshift $z\sim 1000$,
is more than one curvature radius away for $\Omega\simlt0.8$.
Interestingly, in the limit $\Omega\rightarrow0$, an object
located at redshift $z$ can be at most $\ln z$ curvature radii away.

\subheader{Growth of Perturbations}

If the matter present in the universe is slightly inhomogeneously distributed
but has negligible vorticity, and the cosmos has negligible gravity wave
content then the metric of (A1) may be modified to be of the form
$$d s^2=a^2(\eta)\left(-(1+2\Phi)d\eta^2+d\chi^2
                       +(1-2\Phi)\sinh^2\chi\,
                        (d\theta^2+\sin^2\theta\,d\phi^2)\right)\qquad \Phi\ll1
\eqno(\rmA15)$$
where $\Phi(\chi,\theta,\phi,\eta)$ is the Newtonian gravitational potential
induced by the inhomogeneities.  Again assuming the matter pressure is
negligible one finds that the evolution of this potential is given by
$$\Phi(\chi,\theta,\phi,\eta)=\Phi_0(\chi,\theta,\phi)\,F(\eta) \qquad
  F(\eta)=5{\sinh^2\eta-3\eta\sinh\eta+4\cosh\eta-4\over(\cosh\eta-1)^3}
\eqno(\rmA16)$$
where the growing mode solution which is regular at $\eta=0$ has been chosen.
We may see that this function $F(\eta)$, the growth factor, is exponentially
suppressed as $\eta \gg 1$.

\subheader{Scalar Spherical Harmonics on $H^3$}

We shall be interested in the harmonics which form a complete basis for
square integrable functions on $H^3$ [3,8]. These harmonics, written as
$Y_{\nu l m}(\chi,\theta,\phi)$, have the following properties.  The harmonics
satisfy the (Helmholtz) wave equation
$$\bigl[\gamma^{ij}\nabla_i\nabla_j+(\nu^2+1)\bigr]
                              Y_{\nu l m}(\chi,\theta,\phi)=0 \qquad \nu\ge0
.\eqno(\rmA17)$$
The harmonics are orthogonal:
$$\int\sqrt{\gamma} d^3x Y_{\nu l m}(\chi,\theta,\phi)
                  Y^{*}_{\nu' l' m'}(\chi,\theta,\phi)
                                      =\delta(\nu-\nu')\delta_{ll'}\delta_{mm'}
.\eqno(\rmA18)$$
This expression also fixes the normalization of the harmonics.
These harmonics may be expressed in more familiar terms as the Helmholtz solid
harmonics on the hyperboloid
$$Y_{\nu l m}(\chi,\theta,\phi)=\Pi^l_{\nu}(\chi)\,Y_{lm}(\theta,\phi)
\eqno(\rmA19)$$
where $Y_{lm}(\theta,\phi)$ is the usual spherical harmonic on $S^2$, and
$\Pi^l_{\nu}$ is the (Mehler -) Fock harmonic.
$$\Pi^l_{\nu}(\chi)=\mid \Gamma[i\nu+l+1]/\Gamma[i\nu]\mid
			\,\sinh^{-1/2}\chi\,
                    \rmP^{-(l+1/2)}_{i\nu-1/2} (\cosh\chi)
\eqno(\rmA20)$$
Here $\rmP^{-l}_n(z)$ is the Legendre function [9] of the first kind, defined
for $z\ge 1$,
$$\rmP^{-l}_n(z)\equiv{1\over \Gamma(1+l)}\bigl({1-z\over1+z}\bigr)^{l/2}
{}_2F_1[-n,n+1;1+l;{1-z \over 2}]
.\eqno(\rmA21)$$
Useful relations of these functions, as well as a short history, may be found
in [9,10].  To generate values of the associated Legendre functions, we may use
the integral definition, valid for the values of $\mu$ and $\nu$ relevant for
our work,
$$\rmP^{\mu}_\nu(\cosh \chi)=\sqrt{2 \over \pi}
{\sinh^\mu \chi \over \Gamma[{1\over 2} - \mu]}
\int_{0}^{\chi} dt
{ \cosh(\nu + {1\over 2})t \over (\cosh \chi - \cosh t)^{\mu + {1 \over 2}} }
.\eqno(\rmA22)$$
Thus we find
$\rmP^{-1/2}_{i\nu-1/2}(\cosh\chi)
=\sqrt{ 2\over\pi \sinh\chi }{ \sin\nu\chi\over\nu }$.
The recursion relation
$$\rmP^\mu_{\nu+1}(z)-\rmP^\mu_{\nu-1}(z)=
  (2\nu+1)\sqrt{z^2-1}\rmP^{\mu-1}_{\nu}(z)
\eqno(\rmA23)$$
may then be used to more easily generate the functions.  (Note that  the
definition of the associated Legendre functions $P^\mu_\nu(x)$ and
$\rmP^\mu_\nu(z)$ in Gradshteyn \& Ryzhik (1980) is reversed.)

Another notation has been used in the literature to represent the Fock
harmonics (Tomita 1982):
$$\Xi_l(\nu,\chi)=\sqrt{2\over\pi}{(-)^{l+1}\over\sqrt{N_l(\nu)}}\,
                \sinh^l\chi\,\bigl({d\over d\cosh\chi}\bigr)^{l+1}\cos(\nu\chi)
\eqno(\rmA24)$$
with normalization such that
$$\int_0^\infty d\chi\,\sinh^2\chi\,\Xi_l(\nu,\chi)\,\Xi_l(\nu',\chi)
                                                           =\delta(\nu-\nu')
.\eqno(\rmA25)$$
Here, $N_l(\nu)=\prod_{j=0}^l(\nu^2+j^2)$.  For integer values of $l$, i.e.
those functions needed here, the harmonics $\Xi_l(\nu,\chi)$ are
equivalent to $\Pi^l_{\nu}(\chi)$.  However, we shall use the Fock harmonics
expressed with the associated Legendre functions in this paper.

\subheader{Mehler-Fock Transform}

Any scalar function, $\Phi(\vec{x})$, on $H^3$ may be decomposed in terms of
Fock harmonics so long as it is a square integrable function, i.e. the integral
$$\int \sqrt{\gamma} d^3x |\Phi(\vec{x})|^2
\eqno(\rmA26)$$
converges. For the metric given by equation (A1), this condition amounts to
requiring that $|\Phi(\chi,\theta,\phi)e^{-\chi}|$ be finite as
$\chi\rightarrow\infty$.  If satisfied, we may express the function
$\Phi(\vec{x})$ as
$$\Phi(\vec{x})=\int d\nu \sum_{l=0}^\infty\sum_{m= -l}^l
               \widetilde \Phi_{lm}(\nu) Y_{\nu l m}(\vec{x})
.\eqno(\rmA27)$$
Denoting the mode coefficients by a tilde
we find
$$\widetilde{\Phi}_{l m}(\nu)=\int\sqrt{\gamma}d^3x\,\Phi(\vec{x})
                                          Y^{*}_{\nu l m}(\vec{x})
.\eqno(\rmA28)$$
This is the Mehler-Fock transform, which takes the place of the
Fourier transform on $H^3$.

\subheader{Power Spectrum and Correlation Functions for Square Integrable
           Functions}

We shall define the correlation function of a square integrable, spherically
symmetric function to be the volume weighted integral of the product of the
functions evaluated at points separated by a distance $d$.  In particular for
the function $W(\vec{x})$, we define the the correlation function to be
$$\eqalign{
\corrfuncSI{W}(d)
=&{\int \sqrt\gamma d^3x\int \sqrt\gamma' d^3x'\,
W(\vec{x})\,W(\vec{x}')\,\delta(d - |\vec{x} - \vec{x}'|) \over
\int \sqrt\gamma d^3x\,\delta(d - |\vec{x}|)}\cr
=& 4 \pi \int_0^\infty d\nu \nu\,\calP_{W}(\nu) {\sin \nu d \over \sinh d}}
\eqno(\rmA29)$$
Here we define the power spectrum as
$$\calP_{W}(\nu) = \nu^{-2}\sum_{l=0}^\infty \sum_{m= -l}^l
|\widetilde W_{lm}(\nu)|^2 \eqno(\rmA30)$$
For a spherically symmetric function the sum is merely a formality, as the only
non-zero contribution to the Fock transform, $\widetilde W_{lm}(\nu)$, is the
$(l,m)=(0,0)$ mode coefficient.

\subheader{Power Spectra}

The inhomogeneities in our universe, such as the gravitational potential,
$\Phi$, may be described in terms of scalar functions.  Now, the Cosmological
Principle states that we expect the universe to be on average homogeneous and
isotropic. Thus, we expect $\Phi$ to be determined by some homogeneous
isotropic distribution. By these considerations one may determine that the
expectation under this distribution of the second moment of the mode
coefficients is of the form
$$\langle\widetilde{\Phi}_{l m}(\nu )\widetilde{\Phi}_{l' m'}^*(\nu')\rangle
=(2\pi)^3  P_\Phi(\nu) \delta(\nu-\nu')
\delta_{ll'} \delta_{mm'}
.\eqno(\rmA31)$$
Here the function $P_\Phi(\nu)$ is the power spectrum which is given by
$$P_{\Phi}(\nu) =
(2 \pi)^{-3} \sum_{l=0}^\infty \sum_{m= -l}^l |\widetilde \Phi_{lm}(\nu)|^2
\eqno(\rmA32)$$
if the distribution is ergodic.  This power spectrum, $P_\Phi(\nu)$, differs
from the power spectrum for square-integrable function, $\calP_\Phi(\nu)$,
defined in eq~(A31). Since a statistically homogeneous function must maintain
the same level of inhomogeneity in all regions of the space it will not be
square-integrable.  The functions $\widetilde{\Phi}_{lm}(\nu)$ will not be
finite but must be considered as distributions.  The spectrum $P_\Phi(\nu)$,
however, will remain finite or at least integrable, while for a statistically
homogeneous function, $\Phi$, $\calP_\Phi$ would diverge.   Of course,
$P_\Phi$, is non-negative but is otherwise arbitrary.  In terms of the power
spectrum the 2-point correlation function of $\Phi$ may be written as
$$\corrfunc{\Phi}(d)
=\langle\Phi(\chi,\theta,\phi)\,\Phi^*(\chi',\theta',\phi')\rangle
=4\pi\int_0^\infty d\nu \nu
P_\Phi(\nu) {\sin\nu d \over\sinh d} \eqno(\rmA33)$$
where $d$ is the geodesic distance between the two points $(\chi,\theta,\phi)$
and $(\chi',\theta',\phi')$ given in equation (\rmA3).

Let us pause to examine equation (\rmA33), which will give us some insight into
the general claim made in this paper, that large scale structure is suppressed
in an open universe. Observe the dependence on the separation distance $d$ by
the correlation function $\corrfunc{\Phi}(d)$. The integrand is suppressed
exponentially for $d \gg 1$, that is, for separations larger than the curvature
radius. We may recall that on a spatially flat space, the integrand of the
correlation function diminishes only as the inverse of the separation distance.
Thus the 2-point correlation function is exponentially suppressed at large
separations in an open universe, and in some way is more strongly suppressed
than in a spatially flat universe.

\subheader{Convolution Theorem}

If we draw functions from an isotropic, homogeneous random distribution
and convolve each of these functions with a spherically symmetric window
function, we obtain a new distribution of functions which is also
homogeneous and isotropic.
Given a function $W$, the
convolution with $\Phi$ is
$$[W*\Phi](\chi,\theta,\phi)\equiv\int \sqrt{\gamma'} d^3 x'
W(d) \Phi(\chi',\theta',\phi') \eqno(\rmA34)$$
where again $d$ is the geodesic distance between the $(\chi,\theta,\phi)$ and
$(\chi',\theta',\phi')$.  In terms of the transform of $W$, the power
spectra of the convolved function is
$$P_{W*\Phi}(\nu)=(2\pi)^3 \calP_W(\nu) P_\Phi(\nu) \eqno(\rmA35)$$
where $P_W$ is the power spectrum of the convolving function.  If the volume
integral of the convolving function is normalized to unity, then the
convolution is simply a weighted average.  For such an average the variance may
only diminish, i.e.
$$\int \sqrt{\gamma} d^3 x\,W(d)=1 \qquad{\rm implies}\qquad
            (2\pi)^3 \calP_W(\nu)\le1
\quad\vee\nu.\eqno(\rmA36)$$
In flat space, equation (\rmA36) would also imply $(2\pi)^3\calP_W(0)=1$ so
that $P_{W*\Phi}(0)=P_\Phi(0)$, which is just another way of saying that the
mean is not effected by a convolution. However on this hyperbolic space one
finds $(2\pi)^3\calP_W(0)\ne1$ so that $P_{W*\Phi}(0)\ne P_\Phi(0)$.  This is
represented in the fact that the lowest, $\nu=0$, Fock harmonic does not
generally represent the mean value of the spherically symmetric function,
except in the limit $\Rcurv\rightarrow\infty$.  This is another reflection of
the lack of large scale power in an open universe which is demonstrated in this
paper.

\vfill\eject

\vfill\eject

\header{CAPTIONS}

\noindent{\bf Figure 1:} For a 3-d Gaussian function in hyperbolic space these
curves gives the cumulative power in modes with wavenumbers less than $k$ as a
function of $k$.  From bottom to top, the displayed curves have values of the
Gaussian width given by $\chi_0=0.1,\,0.3,\,0.5,\,1,\,2,\,3,\,10$.  The
Gaussians have been normalized to give unit total power.  Observe that if the
Gaussian width is much less than the curvature radius, i.e. $\chi_0\ll1$, that
most of the power comes from $k\sim1/(\chi_0\Rcurv)$, while if $\chi_0\gg1$ the
power comes predominantly from $k$ very close to $1/\Rcurv$.

\noindent{\bf Figure 2:} The ratio of the correlation function of a Gaussian at
spatial
separation $d$ to the zero-lag correlation function is displayed as a function
of $d$. From left to right, the displayed curves have values of
$\chi_0=0.2,\,0.5,\,1,\,2,\,5,\,10$ for the width of the Gaussian.  Observe
that even for a Gaussian with width much greater than the curvature radius,
i.e. $\chi_0\gg1$, that the correlation function drops off rapidly for
$d\simgt1$. Thus, the correlation length for such a Gaussian is never much
greater than the radius of curvature no matter how ``wide'' the Gaussian may
be.

\noindent{\bf Figure 3:} For anisotropies produced on the
surface-of-last-scattering,
this curve gives the upper bound on the ratio of the absolute value of the
correlation function at angular separation $\psi$ to the zero-lag correlation,
as a function of $\psi$ and for various values of $\Omega_0$. From top to
bottom the curves are for $\Omega_0=1,0.75,0.5,0.25,0.1,0.01$. The redshift of
last  scattering used in these figures is $z_{ls} = 1000$. Note that the
required fall-off in the correlation function is very striking even for
moderately small $\Omega_0$.

\noindent{\bf Figure 4:}  The upper and lower bounds on the logarithm of the
ratio of
the zero-lag correlation function smoothed by a Gaussian of widths $7$ and $90$
degrees, as a function of $\Omega_0$ is displayed. For comparison, the central
value of ratio obtained from the COBE 2-year data is displayed has a horizontal
line.  If the anisotropies were completely generated on the surface of
last-scattering and there were no sampling or instrumental errors on this COBE
result then one would conclude that $\Omega_0\simgt 0.05$.

\noindent{\bf Figure 5:} As in figure~3 except that this bound is for purely
adiabatic
fluctuations and the curves, from top to bottom, are for $\Omega_0$ decreasing
from $0.1$ to $0.01$ in units of $0.01$. Observe that the required  fall-off in
the correlation function is less striking than in figure~3 even for small
values of $\Omega_0$.

\noindent {\bf Figure 6}: In a universe where half of spacetime is one
temperature
and half another, this curve gives the probability of observing a hot or cold
circular disc on the sky of angular radius $\theta_\rmd$ or smaller. From
bottom to top the curves correspond to
$\Omega_0=0.99,\,0.5,\,0.2,\,0.1,\,0.05,\,0.02$ in a dust dominated universe,
or a horizon size $\etaobs=0.20,\,1.8,\,2.9,\,3.6,\,4.4,\,5.3$ times the
curvature radius.  Note that if $\Omega_0\ll1$ most observers who observe
hot/cold spots will observe spots with very small angular size.

\end